\documentclass[longauth]{aa}  
\usepackage{graphicx}
\usepackage{txfonts}

\usepackage[pdfpagelabels=false]{hyperref}      
\hypersetup{colorlinks=true,linkcolor=blue,citecolor=blue,filecolor=blue,urlcolor=blue}

\newcommand{\kms}{\rm km\,s$^{-1}$}
\def \kms {{\rm km\,s$^{-1}$}}

\def \micron{\hbox{$\mu$m}}

\def \sol {{\rm M$_\odot$}}

\begin{document} 

\title{Disruption of a massive molecular cloud by a supernova in the Galactic Centre}
\subtitle{Initial results from the ACES project}

\authorrunning{Nonhebel, Barnes, Immer et al. }

\author{M.~Nonhebel\inst{\ref{eso}, \ref{standrews}} \and 
        A.~T.~Barnes\inst{\ref{eso}} \and 
        K.~Immer\inst{\ref{eso}} \and 
        J. Armijos-Abenda\~no \inst{\ref{cardiff},\ref{oaq}} \and
        J. Bally \inst{\ref{boulder}} \and
        C. Battersby \inst{\ref{uconn}} \and
        M.~G.~Burton \inst{\ref{AOP}} \and
        N. Butterfield \inst{\ref{nrao}} \and
        L.~Colzi\inst{\ref{cab}} \and
        P.~Garc\'ia \inst{\ref{UCN},\ref{NAOC}} \and
        A.~Ginsburg \inst{\ref{florida}} \and
        J.~D.~Henshaw \inst{\ref{ljmu}, \ref{mpia}} \and
        Y.~Hu\inst{\ref{ias}} \and
        I.~Jim\'enez-Serra \inst{\ref{cab}} \and
        R.~S.~Klessen \inst{\ref{zah}, \ref{iwr}} \and
        J.~M.~D.~Kruijssen \inst{\ref{tum}, \ref{cool}} \and
        F.-H.~Liang \inst{\ref{eso}, \ref{oxford}} \and
        S.~N.~Longmore  \inst{\ref{ljmu}} \and
        X.~Lu \inst{\ref{SHAO}} \and
        S.~Mart\'in \inst{\ref{esochile}, \ref{jao}} \and
        E.~A.~C.~Mills \inst{\ref{kansas}} \and
        F.~Nogueras-Lara \inst{\ref{eso}} \and
        M.~A.~Petkova \inst{\ref{chalmers}} \and
        J.~E.~Pineda \inst{\ref{mpe}}
        \and
        V.~M.~Rivilla\inst{\ref{cab}} \and
        \'A.~S\'anchez-Monge \inst{\ref{ice}, \ref{ieec}} \and
        M.~G.~Santa-Maria \inst{\ref{florida}} \and
        H.~A.~Smith \inst{\ref{cfa}} \and 
        Y.~Sofue \inst{\ref{UTokyo}} \and        
        M.~C.~Sormani \inst{\ref{insubria}} \and        
        V.~Tolls \inst{\ref{cfa}} \and
        D.~L.~Walker\inst{\ref{jbca}} \and
        J.~Wallace \inst{\ref{uconn}} \and
        Q.~D.~Wang\inst{\ref{umass}} \and
        G.~M.~Williams \inst{\ref{aber}} \and 
        F.-W.~Xu \inst{\ref{koelnph1}, \ref{pkukiaa}, \ref{pkudoa}}}

\institute{
\label{eso} European Southern Observatory (ESO), Karl-Schwarzschild-Stra{\ss}e 2, 85748 Garching, Germany  \and
\label{standrews} SUPA, School of Physics and Astronomy, University of St. Andrews, North Haugh, St. Andrews KY16 9SS, UK \and
\label{cardiff} School of Physics and Astronomy, Cardiff University, The Parade, Cardiff, CF24 3AA, UK \and
\label{oaq} Observatorio Astron\'omico de Quito, Escuela Polit\'ecnica Nacional, Interior del Parque La Alameda, 170136, Quito, Ecuador
\and
\label{boulder}  Department of Astrophysical and Planetary Sciences, University of Colorado, Boulder, CO 80389 USA \and
\label{uconn}{University of Connecticut, Department of Physics, 196A Hillside Road, Unit 3046
Storrs, CT 06269-3046, USA} \and
\label{AOP} Armagh Observatory and Planetarium, College Hill, Armagh, BT61 9DB Northern Ireland\and
\label{nrao} National Radio Astronomy Observatory, 520 Edgemont Road, Charlottesville, VA 22903, USA \and
\label{cab} Centro de Astrobiología (CAB), CSIC-INTA, Carretera de Ajalvir km 4, Torrejón de Ardoz, 28850 Madrid, Spain \and
\label{UCN}{Instituto de Astronom\'ia, Universidad Cat\'olica del Norte, Av. Angamos 0610, Antofagasta, Chile} \and
\label{NAOC}{Chinese Academy of Sciences South America Center for Astronomy, National Astronomical Observatories, CAS, Beijing 100101, China} \and
\label{florida}{Department of Astronomy, University of Florida, P.O. Box 112055, Gainesville, FL 32611, USA.}
\and
\label{ljmu} Astrophysics Research Institute, Liverpool John Moores University, IC2, Liverpool Science Park, 146 Brownlow Hill, Liverpool L3 5RF, UK \and         
\label{mpia} Max Planck Institute for Astronomy, K\"{o}nigstuhl 17, D-69117 Heidelberg, Germany \and
\label{ias} Institute for Advanced Study, 1 Einstein Drive, Princeton, NJ 08540, USA; NASA Hubble Fellow  \and
\label{zah} Universit\"{a}t Heidelberg, Zentrum f\"{u}r Astronomie, Institut f\"{u}r Theoretische Astrophysik, Albert-Ueberle-Stra{\ss}e 2, D-69120 Heidelberg, Germany \and
\label{iwr} Universit\"{a}t Heidelberg, Interdisziplin\"{a}res Zentrum f\"{u}r Wissenschaftliches Rechnen, Im Neuenheimer Feld 205, D-69120 Heidelberg, Germany \and
\label{tum} Technical University of Munich, School of Engineering and Design, Department of Aerospace and Geodesy, Chair of Remote Sensing Technology, Arcisstr. 21, 80333 Munich, Germany \and
\label{cool} Cosmic Origins Of Life (COOL) Research DAO, \href{https://coolresearch.io}{https://coolresearch.io} \and
\label{oxford} Sub-department of Astrophysics, Department of Physics, University of Oxford, Denys Wilkinson Building, Keble Road, Oxford, OX1~3RH, UK \and
\label{SHAO}{Shanghai Astronomical Observatory, Chinese Academy of Sciences, 80 Nandan Road, Shanghai 200030, China} \and
\label{esochile}European Southern Observatory, Alonso de C\'ordova, 3107, Vitacura, Santiago 763-0355, Chile \and
\label{jao} Joint ALMA Observatory, Alonso de C\'ordova, 3107, Vitacura, Santiago 763-0355, Chile \and
\label{kansas} Department of Physics and Astronomy, University of Kansas, 1251 Wescoe Hall Drive, Lawrence, KS 66045, USA \and
\label{chalmers}{Dept. of Space, Earth \& Environment, Chalmers University of Technology, SE-412 96 Gothenburg, Sweden} \and
\label{mpe}{Max-Planck-Institut f\"ur extraterrestrische Physik, Giessenbachstrasse 1, D-85748 Garching, Germany}
\label{ice} Institut de Ci\`encies de l'Espai (ICE, CSIC), Campus UAB, Carrer de Can Magrans s/n, 08193, Bellaterra (Barcelona), Spain \and
\label{ieec} Institut d'Estudis Espacials de Catalunya (IEEC), 08860 Castelldefels (Barcelona), Spain \and
\label{cfa}{Center for Astrophysics Harvard and Smithsonian, Cambridge, MA, 02138, USA} \and
\label{UTokyo}{Institute of Astronomy, The University of Tokyo, Mitaka, Tokyo 181-0015, Japan} \and
\label{insubria}{Universit{\`a} dell’Insubria, via Valleggio 11, 22100 Como, Italy} \and
\label{jbca} UK ALMA Regional Centre Node, Jodrell Bank Centre for Astrophysics, The University of Manchester, Manchester M13 9PL, UK \and
\label{umass}{Department of Astronomy, University of Massachusetts, Amherst, MA 01003, USA} \and
\label{aber}{Department of Physics, Aberystwyth University, Ceredigion, Cymru, SY23 3BZ, UK} \and
\label{koelnph1} I. Physikalisches Institut, Universität zu Köln, Zülpicher Str. 77, D-50937 Köln, Germany  \and
\label{pkukiaa} Kavli Institute for Astronomy and Astrophysics, Peking University, Beijing 100871, People's Republic of China \and
\label{pkudoa} Department of Astronomy, School of Physics, Peking University, Beijing, 100871, People's Republic of China
}


\date{Received -; accepted -}

\abstract{
The Milky Way's Central Molecular Zone (CMZ) differs dramatically from our local solar neighbourhood, both in the extreme interstellar medium conditions it exhibits (e.g. high gas, stellar, and feedback density) and in the strong dynamics at play (e.g. due to shear and gas influx along the bar). 
Consequently, it is likely that there are large-scale physical structures within the CMZ that cannot form elsewhere in the Milky Way. 
In this paper, we present new results from the Atacama Large Millimeter/submillimeter Array (ALMA) large programme ACES (ALMA CMZ Exploration Survey) and conduct a multi-wavelength and kinematic analysis to determine the origin of the M0.8$-$0.2 ring, a molecular cloud with a distinct ring-like morphology. We estimate the projected inner and outer radii of the M0.8$-$0.2 ring to be 79$\arcsec$ and 154$\arcsec$, respectively (3.1 pc and 6.1 pc at an assumed Galactic Centre distance of 8.2 kpc) and calculate a mean gas density $>\,10^{4}$\,cm$^{-3}$, a mass of $\sim$\,$10^6$\,\sol,\ and an expansion speed of $\sim$\,20\,\kms, resulting in a high estimated kinetic energy ($>\,10^{51}$\,erg) and momentum ($>\,10^7$\,$\textup{M}_{\odot}$\,\kms). We discuss several possible causes for the existence and expansion of the structure, including stellar feedback and large-scale dynamics. We propose that the most likely cause of the M0.8$-$0.2 ring is a single high-energy hypernova explosion. To viably explain the observed morphology and kinematics, such an explosion would need to have taken place inside a dense, very massive molecular cloud, the remnants of which we now see as the M0.8$-$0.2 ring. In this case, the structure provides an extreme example of how supernovae can affect molecular clouds.}
  

   \maketitle
%
\section{Introduction}
\label{sec_int}

\begin{figure*}
    \centering
        \includegraphics[width=\textwidth]{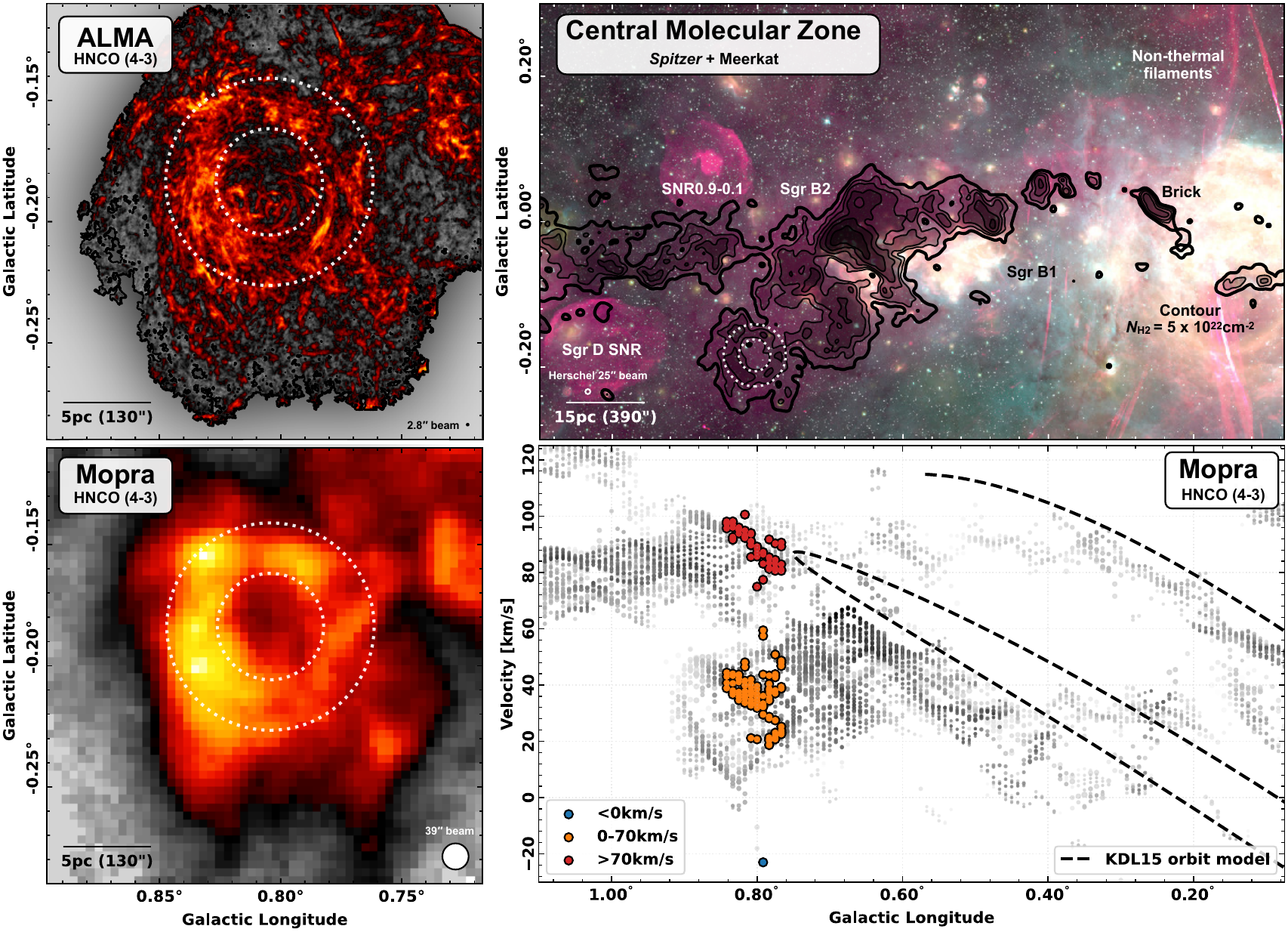}
    \caption{{Overview of the M0.8$-$0.2 ring within the CMZ.} \textit{Upper-left panel:} HNCO\,(4--3) peak intensity map (12 m, 7 m, and total power array data combined) from ACES towards the M0.8$-$0.2 ring. The colour scale extends from 0.2 to 8.0\,K in square root stretch. The dotted white circles show the approximate extent of the \textit{Herschel} 250 $\mu$m continuum emission of the ring-like structure. The ALMA beam size of 2.8\arcsec\ is shown in the lower right, and a scale bar of 5 pc is shown in the lower left of the panel. \textit{Lower-left panel:} HNCO\,(4--3) emission from the Mopra telescope \citep{Jones2012}. The Mopra beam size of 39\arcsec\ is shown in the lower right of the panel. \textit{Upper-right panel:} Multi-wavelength view of part of the CMZ. We show a three-colour composite of 8\,\micron\ emission  from the {\em Spitzer} GLIMPSE survey \citep[green;][]{Churchwell2009}, 24\,\micron\ emission from the {\em Spitzer} MIPSGAL survey \citep[yellow;][]{Carey2009}, and 20\,cm emission observed by MeerKAT \citep[red;][]{Heywood2019, Heywood2022} and the GBT (red; \citealp{Law2008}). Overlaid as black contours are the total molecular gas column densities at levels of 5, 7.5, 10, 15, 25, and 50 $\times\, 10^{22}$\,cm$^{-2}$ (Battersby et al., in prep) calculated using data from the {\em Herschel} Hi-GAL survey \citep{Molinari2010}. Also overlaid are a number of interesting features in the Galactic Centre and the dotted white circles highlighting the ring-like structure (identical to the left panels). The \textit{Herschel} beam size of 25\arcsec\ and a scale bar of 15 pc are shown in the lower left of the panel. \textit{Lower-right panel:} PV diagram of the Mopra HNCO\,(4--3) emission \citep{Henshaw2016a}. Each point corresponds to the longitude and centroid velocity of a Gaussian component of the HNCO emission, extracted using {\sc scouse} \citep{Henshaw2016a}. The emission at the position of the ring-like structure is highlighted with three different colours, corresponding to the velocity ranges <0\,\kms\ (blue), 0--70\,\kms\ (orange), and >70\,\kms\ (red). The orange dots correspond to the M0.8$-$0.2 ring, while the blue and red dots relate to additional line-of-sight material. The orbital model of \citet{Kruijssen2015} is overlaid as a dashed black line.}
    \label{fig_CMZ_HNCO_peak}
\end{figure*}

The Central Molecular Zone (CMZ) is defined as the region within a galactocentric radius of approximately 300 pc (see \citealp{Henshaw2023}). This unique region of the Milky Way hosts the nearest supermassive black hole (e.g. \citealp{GRAVITYCollaboration2018, GRAVITYCollaboration2019}), some of the most massive young star clusters (Arches and Quintuplet; e.g. \citealp{Clark2018}), the largest number of supernovae (SNe) per unit volume (e.g. \citealp{Ponti2015}), and the largest concentration of dense molecular gas in the Galaxy (see Fig.~\ref{fig_CMZ_HNCO_peak}, upper-right panel), making the CMZ an environment unparalleled in the Milky Way.
The interstellar medium (ISM) conditions at the centre of the Galaxy starkly contrast those of the solar neighbourhood. Turbulent motions, magnetic field strengths, gas densities, pressures, and temperatures are orders of magnitude greater than those measured locally (see the recent reviews by e.g. \citealp{Morris1996, Bryant2021, Henshaw2023} and references therein). Many of these properties resemble conditions found at earlier epochs in the Universe \citep{Kruijssen2013}.
In terms of large-scale dynamical effects, the CMZ is also extreme. Inflowing gas rapidly travels along the bar and interacts with pre-existing gas within the CMZ \citep{Sormani2019, Hatchfield2021}, with strong cloud--cloud collisions thought to shape the unique structures of various molecular clouds in the region \citep{Gramze2023}. Furthermore, the CMZ experiences strong gravitational shear due to differential rotation, with powerful tidal forces capable of stretching and tearing apart molecular clouds \citep{Kruijssen2019a}.
Taken together, the extreme environment in terms of its constituents, properties, and dynamics renders the CMZ ripe for structures not observed elsewhere in the Milky Way.

One prominent feature in the CMZ that has garnered significant interest is the M0.8$-$0.2 ring: a ring-like gaseous structure situated in the south-eastern extension of the Sgr\,B2 molecular cloud complex (see Fig. \ref{fig_CMZ_HNCO_peak}, upper-right panel). 
It was initially detected in continuum observations with SCUBA (\citealp{Holland1999}) at 450 and 850~$\mu$m, at resolutions of 8 and 15\arcsec, respectively (PPR G0.80$-$0.18 in \citealp{Pierce-Price2000}). \citealp{Pierce-Price2000} suggested it was a wind-blown region or a supernova remnant (SNR). 
The ring-like structure is also distinctly evident in molecular line emission (Shell 5 in \citealp{Tsuboi2015}, GCM0.83$-$0.18 in \citealp{Mills2017}).
Using H$^{13}$CO$^+$ (1--0) and SiO (2--1) line observations at $\sim40\arcsec$ acquired with the 45 m Nobeyama telescope, \cite{Tsuboi2015} identified the structure as a shell with a diameter of $\sim15$~pc, along with several other shells in the region. 
They interpret the broad velocity-width of the structure in $l-v$ space as evidence of its expansion, and attribute this expansion to successive SN explosions.
Polarimetry studies of the M0.8$-$0.2 ring using the SOFIA telescope indicate that the magnetic field is curved and follows the contours of the emission towards the region \citep{Butterfield2023, Butterfield2024}. 
Additionally, the equipartition magnetic field strength from MeerKAT observations has been estimated to be approximately 0.05 mG \citep{Yusef-Zadeh2022}. \cite{Butterfield2024} similarly argue that this ring-like structure is expanding, compressing the ambient gas and thereby strengthening the magnetic field. 

Numerous studies have highlighted the prevalence of interesting chemistry in the region. The Mopra 3 mm and 7 mm CMZ surveys \citep[e.g.][]{Jones2012, Jones2013} reveal that the M0.8$-$0.2 ring is prominently seen in HNCO, HC$_{3}$N, CH$_{3}$CN, CH$_{3}$OH, and SiO emission, as well as to a lesser extent in CS and N$_{2}$H$^{+}$; however, it is not present in HCN or HNC observations. 
The lines in which it is observed are generally considered dense core or hot core tracers, with HNCO and SiO also known to trace shocks. 
HNCO notably stood out in the fourth component of the principal component analysis (PCA) within the M0.8$-$0.2 ring and in Sgr B2-North (see \citealp{Jones2012}) but nowhere else across the PCA4 map, hinting at some similarity in physical conditions between these two regions. 
SiO was anti-correlated with HNCO in PCA4, possibly indicating slow shocks.
The multi-transition HNCO study by \citet{Martin2008} found a high abundance of HNCO in the region ($\log(X_\mathrm{HNCO}) \approx -8$), again suggesting the presence of shocked gas. 
Together, these findings point 
towards a relatively gentle expansion of the ring-like structure.

In this study we made use of new data from the ALMA CMZ Exploration Survey (ACES), which provides an order of magnitude higher spatial and spectral 
 resolution view of the dense and shocked gas across the whole CMZ. 
In Sect. \ref{sec_obs} we provide an overview of the observations. 
In Sect. \ref{sec_structure} we introduce the structure of the M0.8$-$0.2 ring with the new ALMA observations (Sects. \ref{subsec_morphology} and \ref{subsec_kinematics}) and 
assess its kinematic properties, providing estimates of the mass and energy (Sect. \ref{subsec_mass}). We also compare to previous multi-wavelength observations (Sect. \ref{subsec_multiwave}). 
In Sect. \ref{sec_cause} we discuss possible causes of the structure, including young (Sect. \ref{subsec_youngfeedback}) and late-stage (Sect. \ref{subsec_supernova}) stellar feedback, before considering alternate scenarios such as cloud--cloud collisions and large-scale dynamics (Sect. \ref{subsec_alternatives}). 
Finally, the main results of this paper are summarised in Sect. \ref{sec_conclusions}. 
In the appendices, we provide additional observational information (Appendix \ref{appendixsec_obs}), investigate the radial profiles of molecular line emission across the region (Appendix \ref{appendixsec_radprof}), and present channel maps for a number of lines (Appendix \ref{appendixsec_chans}).

\begin{table}
\caption{Properties of the M0.8$-$0.2 ring.}   
\label{tab:bubble_sum}                                  
\centering                                    
\begin{tabular}{l c }                         
\hline\hline                                  
Property & Value\\
\hline                                  

Longitude, $l$ [deg] & 0.804\\
Latitude, $b$ [deg] & -0.184\\
Radius & \\
\dots $r_\mathrm{in}$/$r_\mathrm{out}$ [arcsec] & 79/154\\
\dots $r_\mathrm{in}$/$r_\mathrm{out}$ [pc] & 3.1/6.1 \\
\dots $r_\mathrm{in,d}$ = $r_\mathrm{in}$/cos($i$) [pc] & 4.4\\
\dots $r_\mathrm{out,d}$ = $r_\mathrm{out}$/cos($i$) [pc] & 8.7\\
Systemic velocity, $v\textsubscript{sys}$ [km/s] & 37.5\\
Expansion speed [km/s] & \\
\dots $v\textsubscript{exp}$  & 15.1\\
\dots $v\textsubscript{exp,d}$ = $v\textsubscript{exp}$/sin($i$) [km/s] & 21.3\\
Age, $t\textsubscript{age}=r_\mathrm{out}/v\textsubscript{exp}$ [Myr] & 0.4 \\
Mass [$\textup{M}_{\odot}$] & \\
\tablefootmark{1} \dots $M$  & 7.6 $\times$\,$10^5$\\
\tablefootmark{2} \dots $M\textsubscript{d}$& 1.1 $\times$\,$10^6$\\
\tablefootmark{3} \dots $M\textsubscript{f}$ & 3.8 $\times$\,$10^5$\\
\tablefootmark{4} \dots $M\textsubscript{d,f}$ & 5.4 $\times$\,$10^5$\\
Kinetic energy [erg] & \\
\dots $E = \frac{1}{2} M v\textsubscript{exp}^2$  & $1.7$ $\times$\,$10^{51}$\\
\dots $E\textsubscript{d} = \frac{1}{2} M\textsubscript{d} v\textsubscript{exp, d}^2$ & $4.8$ $\times$\,$10^{51}$\\
\dots $E\textsubscript{f} = \frac{1}{2} M\textsubscript{f} v\textsubscript{exp}^2$ & $8.5$ $\times$\,$10^{50}$\\
\dots $E\textsubscript{d,f} = \frac{1}{2} M\textsubscript{d,f} v\textsubscript{exp, d}^2$  & $2.4$ $\times$\,$10^{51}$\\
Momentum [$\textup{M}_{\odot}$\,\kms] & \\
\dots $p = M v\textsubscript{exp}$  & 1.1 $\times$\,$10^7$ \\
\dots $p\textsubscript{d} = M\textsubscript{d} v\textsubscript{exp, d}$ & 2.3 $\times$\,$10^7$ \\
\dots $p\textsubscript{f} = M\textsubscript{f} v\textsubscript{exp}$ & 5.7 $\times$\,$10^7$ \\
\dots $p\textsubscript{d,f} = M\textsubscript{d,f} v\textsubscript{exp, d}$ & 1.1 $\times$\,$10^7$ \\
Magnetic Field, $B_\mathrm{POS}$ [mG] & 0.22\,$-$\,0.57 \\
\hline                                       
\end{tabular}

 \tablefoot{We list properties of the M0.8$-$0.2 ring, obtained in Sects. \ref{subsec_morphology}, \ref{subsec_kinematics}, and \ref{subsec_mass}, as well as the plane-of-sky magnetic field strength measured by \citet{Butterfield2024}. A number of different estimates for the radius, expansion velocity, mass, energy and momentum are provided, assuming different possible three-dimensional geometries (see Sect. \ref{subsec_mass}). All `deprojected' values, denoted by subscript $d$, use an inclination angle of $i=45\degr$ (where $\sin(45\degr)=\cos(45\degr)=0.7$) to adjust for an inclined geometry. All mass values are calculated assuming a constant density of $10^{4.2}$\,cm$^{-3}$, determined from radiative transfer modelling. 
 \tablefoottext{1}{Mass calculated assuming a complete spherical shell geometry (i.e. a volume of $v_\mathrm{out}-v_\mathrm{in}= \frac{4}{3} \pi (r_\mathrm{out}^3 - r_\mathrm{in}^3)$).} 
 \tablefoottext{2}{Mass calculated assuming a complete, inclined ellipsoidal shell geometry (i.e. a volume $v_\mathrm{out}-v_\mathrm{in}= \frac{4}{3} \pi (r_\mathrm{out}^2 r_\mathrm{out,d} - r_\mathrm{in}^2 r_\mathrm{in,d}))$.}
 \tablefoottext{3}{Mass calculated assuming a partial, spherical shell geometry, with a filling factor, $f$, of 0.5.}
 \tablefoottext{4}{Mass calculated assuming a partial, inclined ellipsoidal shell geometry, with a filling factor, $f$, of 0.5.}
 }
\end{table}

\section{Observations}
\label{sec_obs}

The observations primarily used in this work were taken from the Atacama Large Millimeter/submillimeter Array (ALMA) cycle 8 large programme `ALMA Central Molecular Zone Exploration Survey' (ACES, 2021.1.00172.L; Longmore et al. in prep), and represent an initial internal release version of the dataset. 
The calibration and imaging were conducted with the ALMA pipeline (\textsc{casa} versions 6.2.1-7 and 6.4.1.12; pipeline versions 2021.2.0.128 and 2022.2.0.64; see \citealp{Hunter2023}), with only minor modifications to account for the size mitigation of the data in the ALMA pipeline and fixing divergent cleans.

To include all spatial scales in the final mosaic, we feathered the (primary beam corrected) 7 m Atacama Compact Array (ACA) and total power cubes in \textsc{casa} \citep{Casa2022}, before feathering these cubes to the (primary beam corrected) 12 m array cube. 
ACES data were imaged onto multiple mosaics, each with a size of $\sim$\,150 pointings, which are then stitched/combined together.
In this case, we made use of the four feathered mosaics covering the area of the M0.8$-$0.2 ring, which are combined to a single cube, using appropriate weighting across the mosaic fields to mitigate for the increased noise at the map edges caused by the primary beam response. 
The complete dataset will be made available to the public with the full description of the data reduction in a later publication (ACES collaboration, in prep.). 
In the case of the 3\,mm continuum, the 12 m array image is feathered to the Green Bank Telescope (GBT) image from MUSTANG Galactic Plane Survey (MGPS90; \citealp{Ginsburg2020}), yielding a circular beam size with a full width at half maximum (FWHM) of 2.52$\arcsec$ and an RMS of $\sim$0.9 mJy beam$^{-1}$. 

The ACES dataset comprises six spectral windows in Band~3 (see Table\,\ref{tab:spectral-line-info} for extracted lines). 
There are two narrow windows with high spectral resolution ($\sim$\,0.2\,\kms channels) around the HNCO(4--3) transition lines, two intermediate spectral resolution ($\sim$\,1.7\,\kms) windows covering H$^{13}$CN(1--0), H$^{13}$CO${+}$(1--0), SiO(2--1), and HN$^{13}$C(1--0), and two broad low spectral resolution ($\sim$\,3\,\kms) windows for continuum observations as well as CS(2--1), SO($3_{2}$--$2_{1}$), and HC$_{3}$N(11--10). 
The working resolution of the HNCO cube is 0.5\,\kms, which provides a balance between resolution and sensitivity (we note that in this work, we are studying the large-scale structure of the region, so the full 0.2\,\kms\, resolution is not required).
The data cubes of the other spectral lines were re-gridded to a spectral resolution of 3\,\kms, for consistency with the lowest spectral resolution window.

Table \ref{tab:spectral-line-info} lists the rest frequencies of the studied transitions, as well as the beam sizes and RMS values of their corresponding cubes. 
The average spatial resolution of the cubes corresponds to a spatial scale of 0.1\,pc.

To gain a multi-wavelength view of the M0.8$-$0.2 ring, we used X-ray observations from \textit{Chandra} \citep{Wang2021}, 2MASS K-band observations \citep{Skrutskie2006}, 
\textit{Spitzer} 8 $\mu$m (GLIMPSE; \citealp{Churchwell2009}, also see \citealp{Ramirez2008, Anand2021a, Arendt2008b}) and 24 $\mu$m (MIPSGAL; \citealp{Carey2009}) observations, \textit{Herschel} 70 to 500 $\mu$m observations \citep{Molinari2010},\footnote{We also examined re-reduced HiGal datasets using the method of \citet{Guzman2015}, which involves a background subtraction technique that more effectively separates Galactic cirrus and other emission from the molecular clouds themselves. We find the differences were not significant and thus retain the HiGal processed results.} Mopra 3 mm observations \citep{Jones2012}, and MeerKAT 20 cm observations \citep{Heywood2022}.

\section{Structure of the M0.8$-$0.2 ring}
\label{sec_structure}

\begin{figure}[!t]
    \centering
        \includegraphics[width=\columnwidth]{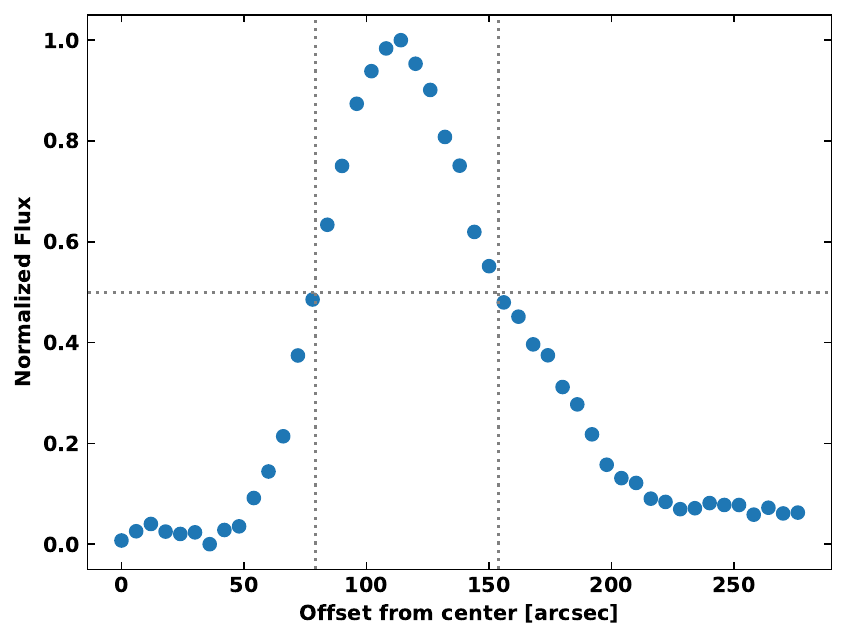}
        \label{fig_chans} \vspace{-3mm}
    \caption{{Radial profile from \textit{Herschel} 250 $\mu$m emission.} The mean flux contained in annuli of increasing radius is plotted, normalised against the peak value. The annuli are defined outwards from the centre of the M0.8-0.2 ring ($l=0.^{\circ}804$ and $b=-0.^{\circ}184$) with a thickness of one pixel. The overlaid vertical grey lines define the radii between which the normalised mean flux is above the threshold of 0.5, indicated by the horizontal grey line.}
    \label{fig_herschel_radial}
\end{figure}

\begin{figure*}[!htp]
    \centering
        \includegraphics[width=\textwidth]{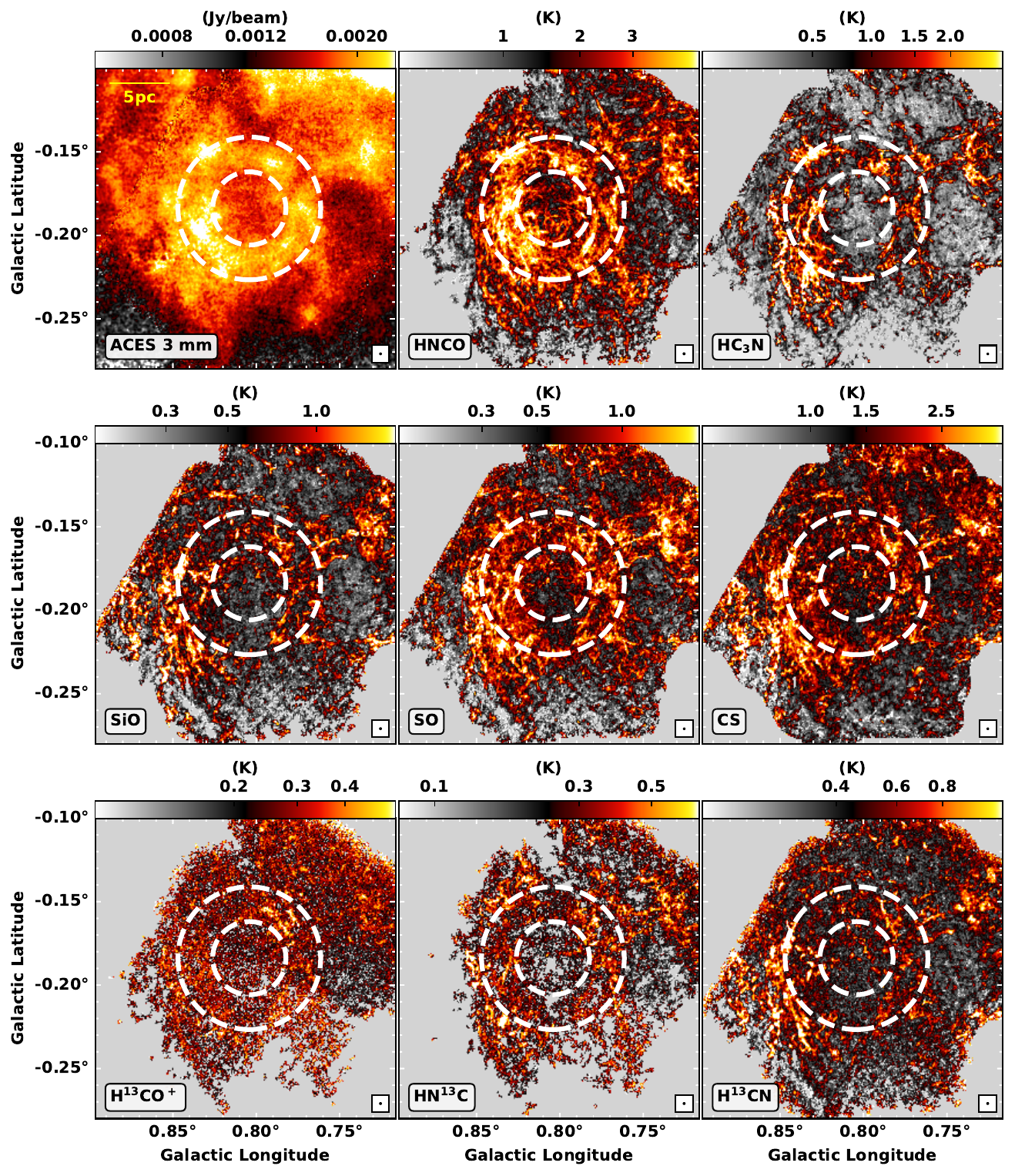}
    \caption{{View of the M0.8$-$0.2 ring from the ACES survey}. The top left image shows the 3 mm continuum emission from the ACES survey (12 m combined with GBT + \textit{Planck} from \citealp{Ginsburg2020}). The remaining panels show the maximum intensity of the spectral transitions in Table \ref{tab:spectral-line-info}, determined within the velocity range $-$5 to 75\,\kms. The cubes were masked with a mask determined by growing a low (2\,RMS) noise threshold mask into a high (5\,RMS) noise threshold mask, and ensuring that this final mask had a continuity of at least one pixel on each side channel. The final masked cubes used to produce the peak intensity images shown will therefore have signal in both the extended lower signal-to-noise pixels and the brightest peaks. The size of the beam of each observation is shown with a circle in the lower-right corner of each image. A scale bar of 5 pc is displayed in yellow on the continuum image. The dashed white circles show the approximate extent of the \textit{Herschel} 250 $\mu$m continuum emission of the ring-like structure.}
    \label{fig_alllines}
\end{figure*}

\subsection{Morphology}
\label{subsec_morphology}

We present high angular resolution HNCO\,(4--3) line emission observations of the M0.8$-$0.2 ring (see Fig.\,\ref{fig_CMZ_HNCO_peak}, upper-left panel). 
Previous observations of this spectral line with the Mopra telescope \citep{Jones2012} are shown in the lower-left panel of Fig. \,\ref{fig_CMZ_HNCO_peak}.
The more than one order of magnitude increase in linear resolution ($\sim$\,200 finer in area) offered by the ACES observations compared to the Mopra data reveals the complex nature of the structure, with a filamentary, `whirlpool-esque' morphology traced by the peak HNCO emission. 

Visually, the majority of the emission from the  molecular cloud falls within a ring surrounding a central depression. 
We used the {\it Herschel} 250\,$\mu$m image (\citealp{Molinari2010}; see the contours in the top right of Fig.\,\ref{fig_CMZ_HNCO_peak} and the middle right of Fig. \,\ref{fig_multiwavelength}) to define the centre and size of the ring-like structure, rather than the ACES continuum image, due to its higher signal-to-noise detection of the large-scale structure and lack of contamination (i.e. the ACES 3\,mm continuum image contains a non-negligible contribution from both free-free and synchrotron emission).
We manually determined the centre of the M0.8$-$0.2 ring to be $l=0.^{\circ}804$ and $b=-0.^{\circ}184$, which is consistent with previous definitions (see e.g. \citealp{Butterfield2024}).
To constrain the inner and outer radii, we used the mean flux radial profile (in steps of one pixel, or 6\,\arcsec) of the {\it Herschel} 250\,$\mu$m emission from the centre of the ring-like structure outwards (see Fig.\,\ref{fig_herschel_radial}). 
There is a well-defined peak corresponding to the observed ring of emission, and we defined inner and outer radii where the mean is 50$\%$ of this peak value.
This yields $r_\mathrm{in}= 79\arcsec$ and $r_\mathrm{out}=154\arcsec$ (Fig.\,\ref{fig_herschel_radial}), which at the distance of the CMZ (8.2\,kpc; \citealp{GRAVITYCollaboration2021}) correspond to 3.2\,pc and 6.2\,pc, respectively. 
In the absence of prior knowledge regarding the three-dimensional structure of the molecular cloud, we also present `deprojected' values assuming an ellipsoid-like geometry that has one axis elongated in the line-of-sight. 
We assumed an inclination angle of $i=45^{\circ}$ (\( \frac{1}{\mathrm{cos}(i)} \approx 1.4 \)), resulting in the major axis of the ellipsoid having $r_\mathrm{in,d}=4.4$\,pc and $r_\mathrm{out,d}=8.7$\,pc.

In Appendix\,\ref{appendixsec_radprof} we produce mean flux radial profiles of the molecular line emission from the ACES dataset studied in this work (Fig. \ref{fig_all_radial}).
We find that the HNCO, HC$_3$N and SO emission profiles are broadly consistent with the continuum definition of the region, but the other lines differ to varying degrees.  
Figure\,\ref{fig_alllines} shows the peak intensity for a number of additional lines from the ACES survey in this region. 
Indeed, the SO emission appears to be most similar to that of HNCO and is concentrated towards the continuum ring, whereas the emission of HC$_{3}$N, CS, SiO, and H$^{13}$CN peaks at larger radii from the centre, possibly tracing denser gas at the edge of the structure (see the critical densities in Table\,\ref{tab:spectral-line-info}). 
It is also interesting to note that the HNCO, HC$_3$N, and SO lines that better trace the ring-like structure are also those with the highest upper energy levels (>10\,K; see Table\,\ref{tab:spectral-line-info}).

\subsection{Kinematics and expansion}
\label{subsec_kinematics}

\begin{figure}
    \centering
        \includegraphics[width=\columnwidth]{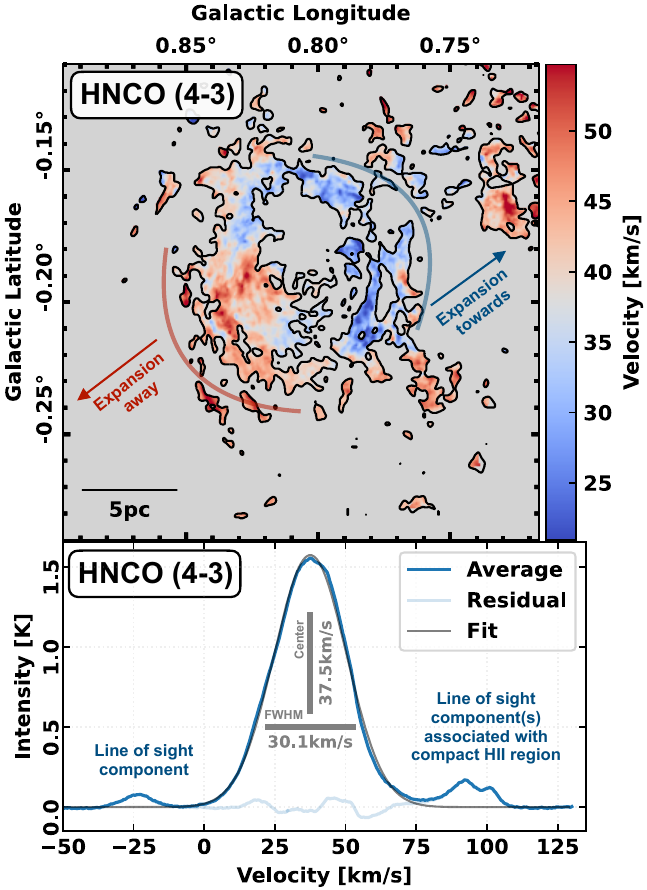}
    \caption{{Overview of the kinematics derived from ACES HNCO\,(4--3) emission across the M0.8$-$0.2 ring.} \textit{Top:} Intensity-weighted centroid velocity integrated between $-$5\,\kms\ and 75\,\kms\ (velocity range of the ring-like structure; see also the channel maps in Fig.\,\ref{fig_chans}). \textit{Bottom:} Spectrum of HNCO\,(4--3) averaged across the ring-like structure (solid blue line). Overlaid as a solid grey line is the single-component Gaussian model fitted to the emission between $-$5\,\kms\ and 75\,\kms, and in faded blue is the residual of the fit. We do not consider the lower ($-$25 \kms) and higher (75 to 110 \kms) velocity components to be associated with the ring-like structure (see Sect. \ref{subsec_morphology}). }
    \label{fig_HNCO_spec_moms}
\end{figure}

A clear velocity gradient is observed across the M0.8$-$0.2 ring in the intensity-weighted centroid velocity (moment 1) image of the HNCO transition (Fig.~\ref{fig_HNCO_spec_moms}; upper panel). As highlighted, the blueshifted (lower) velocities towards the north-west (upper-right) end of this structure indicate that here the gas is moving towards us relative to the systemic velocity, whereas the redshifted (higher) velocities towards the south-east (bottom-left) end show the gas to be receding relative to the systemic velocity. 
The average HNCO spectrum across this region includes a strong broad velocity component, with a FWHM of $\sim$\,30\,\kms, around the systemic velocity of 38\,\kms\ (Fig.~\ref{fig_HNCO_spec_moms}; lower panel) as well as two additional velocity components at $\sim$\,$-$25\,\kms\ and $\sim$\,100\,\kms. 
To understand whether these additional components are related to the ring-like structure, we located this region in the HNCO position--velocity (PV) plots of \citet{Henshaw2016a}. 
Figure \ref{fig_CMZ_HNCO_peak} (lower-right panel) shows the PV plot of \citet{Henshaw2016a} for the same area of the CMZ as the upper-right panel in the same figure.
The M0.8$-$0.2 ring is highlighted with three colours for the three velocity components. 
The three components are clearly distinct, with the higher velocity component being associated with the material that bridges to the 1.3 degree cloud complex at larger longitudes, and an extension of the turning-point of the orbital model from \citet{Kruijssen2015}. 
Additionally, the HNCO emission from the velocity components at $-$25 and 100\,\kms\ from the ACES observations do not show any morphological similarity with the HNCO emission of the main component (see the channel maps in Fig.\,\ref{fig_HNCO_chans}). 
We thus concluded that these two components represent line-of-sight material unassociated with the M0.8-0.2 ring and hence omitted them from our analysis.

We fitted a single Gaussian profile to the 38\,\kms\ component present in the HNCO emission (Fig.~\ref{fig_HNCO_spec_moms}; upper panel) using the {\sc PySpecKit} package \citep{Ginsburg2011,Ginsburg2022}. 
We determined a centroid velocity ($v\textsubscript{cent}$) of 37.5$\pm$1.7\,\kms and a line-width ($v\textsubscript{FWHM}$) of 30.1$\pm$4.0\,\kms\ (after subtraction of the contribution of the channel width). 
One interpretation of this velocity profile is that it is the result of the large-scale expansion of the structure (see \S\,\ref{subsec_alternatives} for alternative scenarios). Here we interpreted the blueshifted and redshifted material to represent the foreground and background parts of a three-dimensional shell of material, respectively, which along the line-of-sight is observed as a ring of emission. 
We estimated the expansion velocity to be $v\textsubscript{exp} = (v\textsubscript{FWHM})/2 = $\,15.1\,\kms. 
\footnote{The full width at tenth maximum (FWTM) could also be employed in this calculation to encompass the more extreme velocities, which would increase the value by a factor of (FWTM/FWHM) $\sim$ 1.8.}
We acknowledge that this estimate is a lower limit due to projection effects and calculated an adjusted expansion velocity for an ellipsoidal structure.
Again, we conservatively assumed an inclination angle of \( i = 45^\circ \), which gives $v\textsubscript{exp, d} \sim 21.3$\,\kms.

These estimates of the expansion velocity differ from that reported by \citet[][$v\textsubscript{exp}\sim$\,40\,\kms]{Tsuboi2015}. 
This is due to two factors: 1) \citet{Tsuboi2015} used lower angular resolution ($\sim$\,$40$\arcsec) longitude-velocity ($l-v$) diagrams to analyse the region, rather than fully resolving the position--position--velocity (PPV) structure as in the ACES observations; 2) the observations from the Nobeyama Radio Observatory used by these authors focus on SiO and H$^{13}$CO$^+$, as opposed to HNCO where the M0.8$-$0.2 ring can be seen more clearly (Fig.\,\ref{fig_multiwavelength}). 
This resulted in the authors also including the additional higher velocity component (at $\sim$\,100\,\kms; see Figs.\,\ref{fig_HNCO_spec_moms} and \ref{fig_HNCO_chans}), leading to an overestimation of the velocity width of the region and, as such, its expansion speed. 

Working on the assumption that the velocity gradient represents expansion, we estimated the age of the M0.8$-$0.2 ring ($t_\mathrm{age} = r_\mathrm{out} / v_\mathrm{exp}$). 
Assuming the velocity has remained constant throughout the expansion, we calculated $t_\mathrm{age} \sim 0.4$ Myr (note that $t_\mathrm{age} = r_\mathrm{out} / v_\mathrm{exp} = r_\mathrm{out,d} / v_\mathrm{exp,d}$).

\subsection{Mass, energy, and momentum}
\label{subsec_mass}

\begin{figure}
    \centering
        \includegraphics[width=\columnwidth]{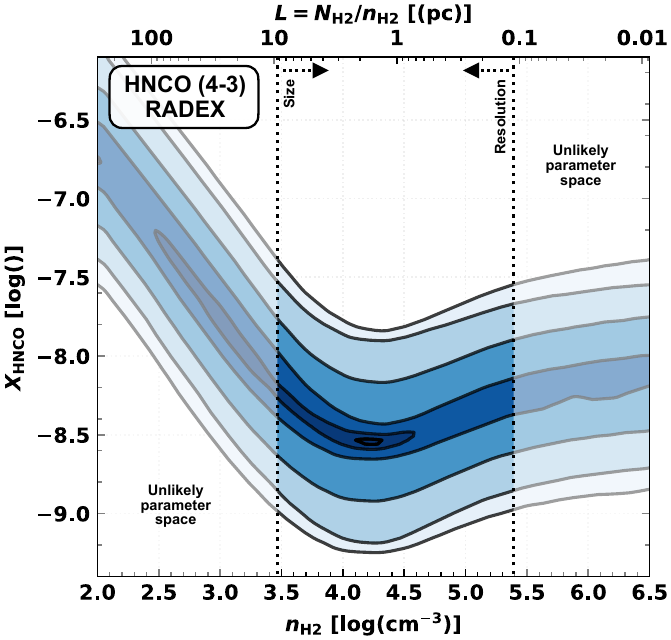}
    \caption{{Radiative transfer modelling of the HNCO\,(4--3) emission.} We demonstrate the likelihood of achieving the observed intensities (1 to 5\,K) from a grid of models. This plot shows contours at the 10, 25, 50, 75, 90, and 95 per cent likelihoods of achieving the observed brightness temperatures, through varying the $T_\mathrm{kin}$, line width, and HNCO abundance. On the upper $y$-axis, we show the predicted depth calculated from the column and number densities ($L = N_{\mathrm{H_2}}/n_{\mathrm{H_2}}$). The vertical dashed lines show the angular resolution of the observations and the measured size of the ring-like structure, which we use as upper and lower bounds to constrain the parameter space. We find a peak likelihood of $n_{\mathrm{H_2}} \approx 10^{4.2}$\,cm$^{-3}$, with a large spread in plausible densities between $10^{3.5}$ and $10^{5.5}$\,cm$^{-3}$.}
    \label{fig_HNCO_radex}
\end{figure}

There are multiple mass estimates for the M0.8$-$0.2 ring. 
For instance, \citet{Mills2017} determined a mass of $2.6 \times 10^5$\,\sol\ (see their Fig.\,3), employing an iterative cirrus fore/background subtraction technique to \textit{Herschel} data (similar to e.g. \citealp{Guzman2015}) to produce a column density of dense gas (Battersby et al. in prep).\footnote{Assuming a gas-to-dust ratio of 100, yet this could be somewhat lower in the Galactic Centre due to its super-solar metallicity (see the discussion in \citealp{Barnes2019}). Moreover, we note that this estimate does not take into account the fore- and background line-of-sight contamination from objects within the Galactic Centre.}
\citet{Tsuboi2015} estimated a mass of $2.7 \times 10^3$\,\sol\ from radiative transfer modelling of CS\,(1--0) emission (although the exact geometry assumed to convert the derived density to mass in this calculation is not clear).
Given the significant discrepancy between these previous mass estimates, we performed a revised calculation based on the new HNCO\,(4--3) emission from ACES using the one-dimensional non-local thermodynamic equilibrium radiative transfer code {\sc Radex} (see Fig.\,\ref{fig_HNCO_radex}; \citealp{vanderTak2007}).

We made use of the {\sc SpectralRadex} python package implementation of {\sc Radex} \citep{Holdship2021} to explore a grid of models by varying the following parameters:
The kinetic temperature ($T_\mathrm{kin}$) ranged from 30 to 150\,K with a step size of 10\,K. 
The number density of molecular hydrogen ($n_\mathrm{H_2}$) spanned $10^2$ to $10^8$ cm$^{-3}$ with a step size of 0.05 dex.
The line-width values considered were 5, 10, and 20\,\kms.\footnote{Note that these do not necessarily correspond to the observed line-width, but rather to one of the dimensions of the radiative transfer box that we expect to be significantly smaller than our mean measured values across the region. These values are more similar to those measured on a per-beam basis across the region (a future work will investigate the full velocity component structure of the region).} 
The abundance of HNCO (\(X_\mathrm{HNCO}\)) was varied from \(10^{-10}\) to \(10^{-6}\) with a step size of 0.05 dex. 
We assumed a fixed molecular hydrogen column density (\(N_\mathrm{H_2}\)) of \(8.5 \times 10^{22}\) cm\(^{-2}\) (i.e. to convert the HNCO abundance to column density), which represents an average measured value from \textit{Herschel} column density observations across the M0.8-0.2 ring.
We note here the consistency of our assumed ranges with the values derived over a single IRAM 30 m pointing on the ring-like structure of \(N_\mathrm{H_2}\) $\approx$ \(7 \times 10^{22}\) cm\(^{-2}\) and $\log(X_\mathrm{HNCO}) \approx -8$, estimated using a multi-transition analysis (\citealp{Martin2008}; note that here we can only make use of a single transition of HNCO covered by the ACES survey).

The results from the grid of models, as illustrated in Fig.\,\ref{fig_HNCO_radex}, depict the frequency with which the models produce brightness temperatures within the observed range of 1 to 5 K for the M0.8$-$0.2 ring (see the upper-left panel of Fig.\,\ref{fig_alllines}). 
Figure\,\ref{fig_HNCO_radex} indicates that the inferred abundance of HNCO remains approximately constant (within the 90 per cent likelihood contour) at $\log(X_\mathrm{HNCO}) \approx -8.5$ above a number density of around $10^4$ cm$^{-3}$. 
Below this density (close to the critical density of emission for HNCO, 2.1\,$\times$\,$10^4$ cm$^{-3}$; see Table\,\ref{tab:spectral-line-info}), the abundance must increase to reproduce the observed brightness of HNCO emission. 

In Fig.\,\ref{fig_HNCO_radex} we also show the predicted depth ($L = N_{\mathrm{H2}}/n_{\mathrm{H2}}$) calculated from the observed column density and the model number density. 
Bounds on the depth are imposed, assuming that the structure should not be much deeper than its projected size ($\sim 10$ pc) and should not be smaller than the angular resolution of our observations ($\sim 0.1$ pc). 
The analysis yields a peak likelihood of $n_{\mathrm{H2}} \approx 10^{4.2}$ cm$^{-3}$,\footnote{At this maximum likelihood density of $n_{\mathrm{H2}} \approx 10^{4.2}$, and with a $\log(X_\mathrm{HNCO}) \approx -8.5$, the excitation temperatures of HNCO range from around 10 to 40\,K, and the optical depth ranges from around 0.1 to 0.5.} with a spread in plausible densities ranging from $10^{3.5}$ to $10^{5.5}$ cm$^{-3}$.
A simple check of this value can be done by comparing our HNCO-derived density with what we get if we divide the mean H$_2$ column density by twice the shell width (assuming we have observed both the front and back sides). 
For a width of 1\,pc and $N_{\mathrm{H2}} = 8.5 \times 10^{22}$ cm$^{-2}$, this gives a density $n_{\mathrm{H2}} \approx 10^{4.1}$ cm$^{-3}$, which is consistent with our value.

We estimated the mass of the M0.8$-$0.2 ring given various possible three-dimensional geometries. Taking a constant density of $10^{4.2}$\,cm$^{-3}$, and assuming a complete spherical shell geometry with inner and outer radii of 3.1\,pc and 6.1\,pc (i.e. with a volume of $v_\mathrm{out}-v_\mathrm{in}= 4/3 \pi (r_\mathrm{out}^3 - r_\mathrm{in}^3))$, we calculated a total mass of $M$\,=\,7.5\,$\times$\,10$^{5}$\,\sol. 
Assuming a complete ellipsoidal shell geometry again inclined at $45^\circ$, with one axis elongated in the line-of-sight and the other two as observed (i.e. with a volume of $v_\mathrm{out}-v_\mathrm{in}= 4/3 \pi (r_\mathrm{out}^2 r_\mathrm{out,d} - r_\mathrm{in}^2 r_\mathrm{in,d}))$, we calculated a mass of $M\textsubscript{d}$\,=\,1.1\,$\times$\,10$^{6}$\,\sol. 
Assuming instead a clumpy or partial shell with a filling factor of a half (i.e. $f=0.5$) decreases both of these mass estimates by a factor of 2.
Our best estimate of the mass is given by adjusting both the inclination and filling factor, yielding $M\textsubscript{d,f}$\,=\,5.4\,$\times$\,10$^{5}$\,\sol.
These estimates are in agreement with the mass determined from the {\it Herschel}-derived H$_2$ column density map \citep{Mills2017}, but a factor of 200 higher than the value of \citet{Tsuboi2015}. 

Using our most realistic mass estimate of a partial, inclined ellipsoidal shell ($M\textsubscript{d,f}$) and taking the inclination-corrected expansion speed ($v\textsubscript{exp,d}$) of the M0.8$-$0.2 ring, we calculated its kinetic energy ($E\textsubscript{d,f} = 0.5 M\textsubscript{d,f} v\textsubscript{exp,d}^2$) and momentum ($p\textsubscript{d,f} = M\textsubscript{d,f} v\textsubscript{exp,d}$).
We find $E\textsubscript{d,f} = 2.4$\,$\times$\,$10^{51}$\,erg and $p\textsubscript{d,f} = 1.1$\,$\times$\,$10^7$\,$\textup{M}_{\odot}$\,\kms (see Table\,\ref{tab:bubble_sum} for a summary of all mass, energy and momentum estimates).

\subsection{Multi-wavelength view}
\label{subsec_multiwave}

\begin{figure*}
    \centering
        \includegraphics[width=\textwidth]{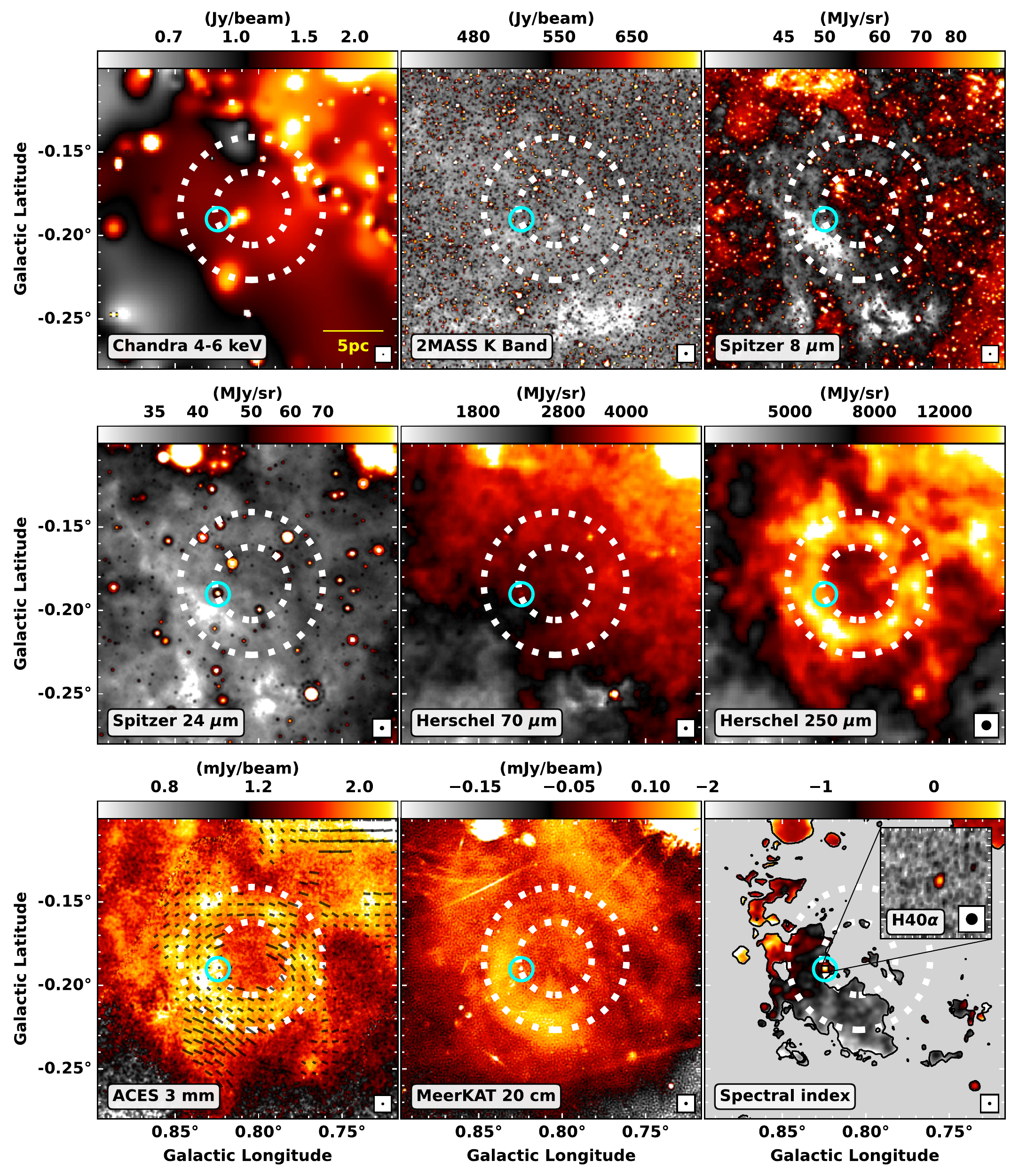} \vspace{-7mm}
    \caption{{Multi-wavelength view of the M0.8$-$0.2 ring.} 
    The dotted white circles in each panel show the approximate extent of the \textit{Herschel} 250 $\mu$m continuum emission of the ring. The cyan circle marks the position of the \ion{H}{ii} region discussed in Sect. \ref{subsec_youngfeedback}.
    \textit{Top row, left to right:} \textit{Chandra} 4 - 6\,keV emission \citep{Wang2021}, 2MASS K-band observations \citep{Skrutskie2006}, and \textit{Spitzer} 8 $\mu$m observations from the GLIMPSE survey \citep{Churchwell2009}. \textit{Middle row, left to right:} \textit{Spitzer} MIPSGAL 24 $\mu$m \citep{Carey2009} and \textit{Herschel}  70 and 250 $\mu$m observations \citep{Molinari2010}. \textit{Bottom row, left to right:} 3\,mm continuum emission from the ACES survey (12\,m combined with GBT+\textit{Planck} from \citealp{Ginsburg2020}), 20\,cm emission observed by MeerKAT, and the MeerKAT spectral index map, with a zoomed-in view of the peak H40$\alpha$ emission at the location of the \ion{H}{ii} region overlaid. 
    In the 3\,mm continuum image, magnetic field pseudo-vectors from the FIREPLACE polarimetry survey are overlaid (polarisation vectors rotated by 90 degrees, as the direction of the polarised electric field measured in the infrared is considered to be perpendicular to that of the magnetic field; \citealp{Butterfield2023, Butterfield2024}).
    The beam sizes of the observations are shown with a circle in the lower-right corner of each panel, with the FWHMs as follows (proceeding from left to right and top to bottom, and finishing with the H40$\alpha$ image): 0.5, 4, 2, 6, 5, 18, 2.5, 4, 4, and 1.4\arcsec. A scale bar of 5\,pc is displayed in yellow on the \textit{Chandra} image.}
    \label{fig_multiwavelength}
\end{figure*}

An extensive multi-wavelength analysis of the region has been performed by \citet{Butterfield2024}, and here we provide a brief overview of their findings (see Fig.\,\ref{fig_multiwavelength}).
There are two X-ray point sources located, in projection, at the centre of the M0.8$-$0.2 ring and one on the southern edge, but no widespread X-ray emission has been detected. 
Two infrared point sources, present in 8 and 24 $\mu$m 
\textit{Spitzer} observations, similarly appear to be located within the ring-like structure \citep{Butterfield2024}.
Following up on this, we checked the infrared data from the VISTA Variables in the Via Lactea (VVV) survey \citep{Minniti2010} and found no overdensity of stars at these locations, or elsewhere within the M0.8$-$0.2 ring, though this does not discount young stars being present in the region (as in Sgr B1 and C, \citealp{Nogueras-Lara2022, Nogueras-Lara2024}).
Additionally, if there is a cluster present, extinction may be too high to identify it \citep[see Fig. 7 in][]{Henshaw2022}. 
Indeed, the infrared observations at 8 and 24 $\mu$m reveal an infrared dark (i.e. extinction) feature along the eastern to southern arc of the ring-like structure that coincides with the strongest dust continuum emission at 250 $\mu$m and 3\,mm. 
This arc also shows strong radio emission in the MeerKAT observations at 20\,cm \citep{Heywood2019, Heywood2022}. 
The infrared and MeerKAT observations reveal a point source on the eastern inner edge of the ring, which is believed to be a compact \ion{H}{ii} region (discussed further in Sect. \ref{subsec_youngfeedback}). 
The M0.8$-$0.2 ring is also well traced by dust polarisation vectors from the FIREPLACE polarimetry survey (\citealp{Butterfield2023, Pare2024}), indicating a clear connection between the magnetic field and the ring-like structure, as explored in detail in \citet{Butterfield2024}. 

\section{Cause of the M0.8$-$0.2 ring}
\label{sec_cause}

We next explored possible explanations for the morphology and velocity profile of the M0.8$-$0.2 ring, taking into consideration the energy and momentum estimates (Sect. \ref{subsec_mass}), along with a range of multi-wavelength ancillary observations (Sect. \ref{subsec_multiwave}). Stellar feedback, both from young stars and SNe, as well as alternative explanations, such as cloud--cloud collisions and large-scale dynamics, were considered. 

\subsection{Feedback from young high-mass star(s)}
\label{subsec_youngfeedback}

\begin{figure}
    \centering
        \includegraphics[width=\columnwidth]{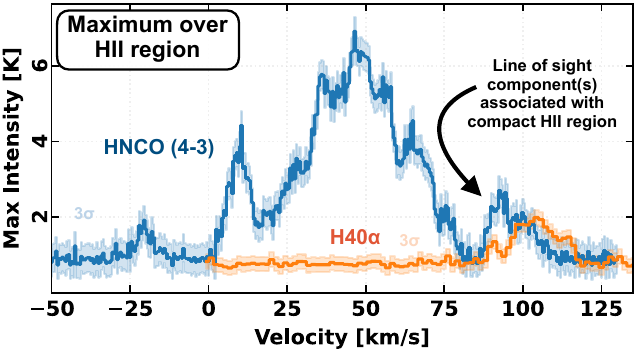}
    \caption{{Molecular and ionised gas kinematics towards the compact \ion{H}{ii} region (marked with a cyan circle in Fig.\,\ref{fig_multiwavelength}).} Shown as the solid blue and orange lines are the maximum ALMA (ACES) HNCO\,(4--3) and H40$\alpha$ specta taken from a $\sim$\,1\,pc (30\arcsec) aperture around the \ion{H}{ii} region (to include both the ionised and associated molecular gas; the same size as the cyan circle in Fig. \ref{fig_multiwavelength}). We show that the higher velocity component ($\sim$\,100\,\kms), which is unassociated with the ring-like structure, is coincident with the ionised gas from the \ion{H}{ii} region (no significant H40$\alpha$ emission is seen within the ring-like structure's velocity range).}
    \label{fig_H40a_spectrum}
\end{figure}

Young high-mass stars are known to inject a large amount of energy and momentum into their surroundings (e.g. \citealp{Krumholz2014, Rosen2020}), which can lead to the expansion of bubbles of material (e.g. \citealp{Barnes2022, Henshaw2022}). 
In this context, of interest is a 3\,mm point source located at $l=0.^{\circ}8244$, $b=-0.^{\circ}1903$ (marked with a cyan circle in Fig.\,\ref{fig_multiwavelength}). 
\citet{Butterfield2024} show that this source is present in observations ranging from the near-infrared to cm wavelengths.
Observations from MeerKAT at $\sim$20\,cm \citep{Heywood2022, Yusef-Zadeh2022} show that the spectral index of the source is approximately +0.7 (see Fig. \ref{fig_multiwavelength}), suggesting the origin of the source is free-free emission from ionised gas, most likely from a compact \ion{H}{ii} region.
The fact that this source is also seen in the infrared is also suggestive that this emission is free-free in nature, as opposed to a synchrotron source.
In addition, we find a point source in H40$\alpha$ emission towards this location (see Figs.\,\ref{fig_multiwavelength} and \ref{fig_H40a_spectrum}), which confirms the identification of this source as a compact \ion{H}{ii} region. 
However, we find that the ACES H40$\alpha$ spectrum peaks at $\sim$100\,\kms, which differs significantly from the molecular gas associated with the M0.8$-$0.2 ring at this position ($\sim$50\,\kms). 
Rather, the velocity of the H40$\alpha$ emission is consistent with the higher velocity component seen at around 100\,\kms\ in the average spectrum of HNCO (see Fig.\,\ref{fig_CMZ_HNCO_peak}), which is morphologically different and does not seem to be associated with the ring-like structure (see Fig.\,\ref{fig_HNCO_chans}).

We further ruled out this source as the progenitor of the M0.8$-$0.2 ring based on mass. 
To calculate its mass, we first calculated the Lyman continuum ionising flux using the following, as set out by \cite{Schmiedeke2016}:

\begin{equation}
N\textsubscript{LyC} = 4.771 \times 10^{42} \left[\dfrac{S_{v}}{\textrm{Jy}}\right] \left[\dfrac{{v}}{\textrm{GHz}}\right]^{0.1}  \left[\dfrac{T\textsubscript{e}}{\textrm{K}}\right]^{-0.45} \left[\dfrac{d}{\textrm{pc}}\right]^2 \text{s}^{-1}, 
\end{equation}
where $S_v$ is the integrated flux density of the source, measured in the 12 m-only continuum map to be 36\,mJy, with a central frequency, $v$, of 98.6\,GHz, 
$T_\textsubscript{e}$ is the electron temperature, assumed to be 5000\,K (typical for Galactic Centre \ion{H}{ii} regions; e.g. \citealp{Mehringer1992}), and $d$ is the distance to the Galactic Centre of 8.2\,kpc \citep{GRAVITYCollaboration2021}. 
Inputting these values, we obtained an ionising Lyman continuum flux of $\log_{10}$(N\textsubscript{LyC}/s$^{-1}$) $\approx$ 47.6. 
This value is consistent with a late O- or early B-type zero-age main-sequence star (spectral type O9V/B0V; \citealp{Panagia1973, Smith2002}) with a mass of $\sim$\,20\,$\textup{M}_{\odot}$ (\citealp{Nieva2014}). 
Despite being massive, this single star is not able to drive the expansion of such a massive (10$^{5-6}$\,\sol) bubble over such a short timescale ($\sim$0.4\,Myr). 
For example, such a star produces $\approx$ $10^{4}$\,L$_{\odot}$ $\times$ 0.4\,Myr $\approx$ 5$\times$10$^{50}$\, erg of energy over the lifetime of the M0.8$-$0.2 ring: an insufficient amount given the energy estimate of the structure (see Sect. \ref{subsec_mass}).
Assuming a conservative coupling efficiency of 10$\%$ and an escape fraction of 50$\%$ (i.e. equal to the filling factor of the ring-like structure), $\sim$\,500 similar O9/B0 stars are needed to power the expansion of the structure (i.e. to achieve $E\textsubscript{d,f}$).
If this were the case, and a significant fraction of the ionising radiation was escaping from the region, we would expect to see some influence on the ionisation state of the gas surrounding the M0.8$-$0.2 ring.
Such a feature may not be visible in the infrared due to high extinction, but would be observed clearly as free-free emission in the radio (e.g. as in the case of Sgr B1; \citealp{Barnes2020b}).

\begin{figure*}
    \centering
        \includegraphics[width=\textwidth]{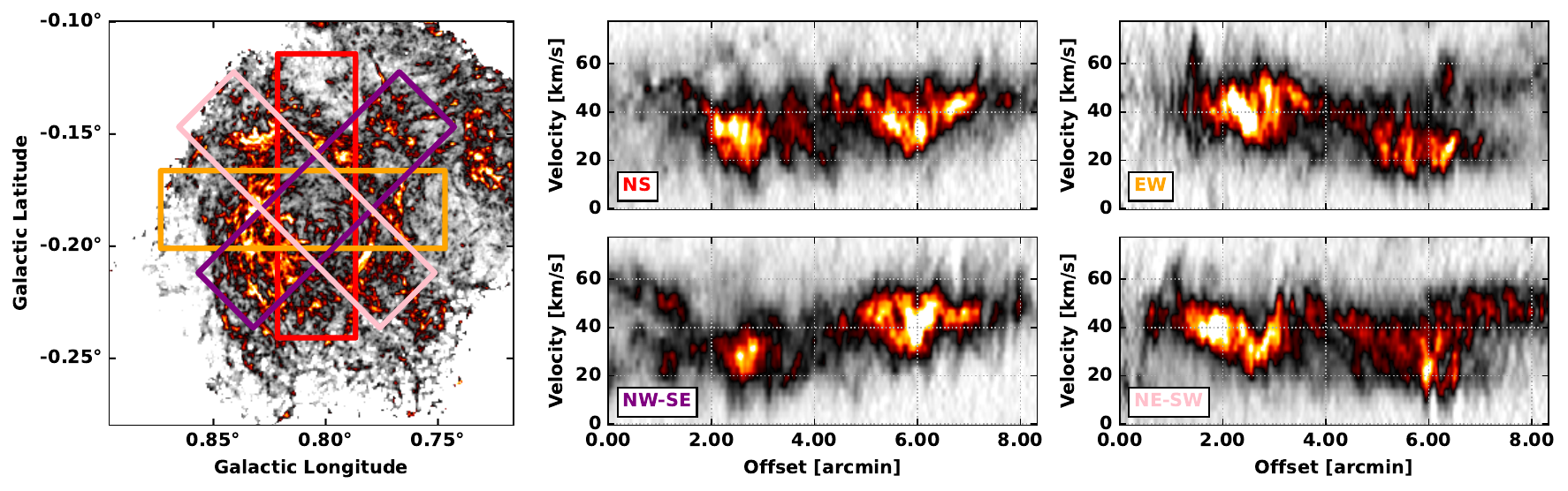}
    \caption{{HNCO PV plots.} \textit{Left panel}: Maximum intensity HNCO(4--3) map from the ACES survey. Overlaid are four rectangles showing the area over which the PV cuts were performed. \textit{Right panels}: PV plots for four cuts through the ring-like structure. The cuts were performed from north to south (NS, red), east to west (EW, orange), north-west to south-east (NW-SE, purple), and north-east to south-west (NE-SW, pink).}
    \label{fig_pv-cuts}
\end{figure*}

\subsection{Feedback from supernova(e)}
\label{subsec_supernova}

Alternatively, the M0.8$-$0.2 ring may be the result of SN feedback \citep{Pierce-Price2000, Tsuboi2015}. 
Indeed, there is an arc of extended, diffuse (20\,cm) radio emission that traces the south-east (lower left) of the ring-like structure (see Fig.\,\ref{fig_multiwavelength}). 
In contrast to the radio point source (Sect. \ref{subsec_youngfeedback}), the associated spectral indices of this extended radio continuum emission are negative, ranging from -0.4 to -1.7 moving south along the emission (Fig. \ref{fig_multiwavelength}). 
Such values point towards the emission being synchrotron emission, as is commonly observed towards SNRs (see \citealp{Anderson2017,Heywood2022}).\footnote{Though we note the possible contribution of non-thermal emission mechanisms towards \ion{H}{ii} regions, as studied in the Sgr B2 region in the CMZ (e.g. \citealp{Meng2019, Padovani2019}).} 
Making use of \textit{Chandra} observations, we find no extended X-ray emission associated with the M0.8$-$0.2 ring (Fig.\,\ref{fig_multiwavelength}). 
The lack of X-ray emission, frequently observed towards young SNRs as a result of shock-heated plasma (\citealp{Vink2017}), and low radio brightness of the source are in line with an evolved SNR. 
The large amount of swept-up mass, typical of SNRs in the later `radiative' or `snowplough' phase (\citealp{Vink2012}), and the observed orientation of the magnetic field, explored by \cite{Butterfield2024}, are also consistent with an older SNR. 
Typically, young SNRs have radial magnetic fields, whereas more evolved remnants display magnetic fields tangential to the shock front \citep{Dubner2015}, as observed here (see Fig.~\ref{fig_multiwavelength}). Given the non-linear expansion of an SNR over time, with an initial rapid expansion followed by a slow increase in radius as it enters the later stages of its evolution, our initial estimate of the age of the structure ($t = r/v = 0.4$\,Myr) will be a factor of a few too large.  

We next compared the kinetic energy of the M0.8$-$0.2 ring to that of a typical SNR. 
A single SN event is thought to inject $\sim$ $10^{51}$ erg of baryonic energy into the ISM (\citealp{Bethe1990,Thornton1998,Kim2015a}). 
A significant amount of this energy is converted into radiative energy over time, with typical coupling efficiencies (i.e. the percentage of kinetic energy retained by the surrounding medium and not radiated away) varying from a few to a few tens of per cent (e.g. \citealp{Thornton1998,Tamburro2009,Sharma2014,Kim2015a,Martizzi2015,Gentry2017}). 
The total energy available to drive large-scale motions of the ISM from a single SN is therefore $\sim10^{49}-10^{50}$ erg. 
Making comparison to our most realistic estimate for the kinetic energy of the M0.8$-$0.2 ring ($\sim$ 2.4\,$\times$$10^{51}$\,erg; see Sect. \ref{sec_structure}), a coupling efficiency of $>100\%$ is required for a single SN to be the driving force behind the expansion of the structure. 
If late-stage stellar feedback is indeed behind the ring-like structure, there are two solutions to this unphysical coupling efficiency: multiple SNe acting in concert, or a single abnormally luminous event (i.e. a hypernova).

Multiple, clustered SNe could provide an explanation as to the very high kinetic energy and momentum of the M0.8$-$0.2 ring. To derive an approximate minimum number of SNe required to power the expansion of the structure, we assumed a generous coupling efficiency of 30$\%$ and an injected energy of $10^{51}$ erg. 
Assuming the impact of multiple SNe is additive, and using the estimated kinetic energy of the M0.8$-$0.2 ring of $\sim$ 2.4\,$\times$$10^{51}$\,erg, we find eight SNe are required. 
We note that the effect of subsequent SN explosions is not well understood, and clustering of SNe, both temporally and spatially, may have a large impact on the energy and momentum injected per SN (see \citealp{Krumholz2019a} for a full discussion). 
Despite this, we can conclude that many concentrated SNe would have to have occurred within a relatively small (around 10 pc) region.  
However, the strong gravitational shear within the Galactic Centre will disperse a stellar cluster as it evolves on a relatively short timescale, thus making numerous spatially coincident SNe unlikely unless a cluster is tightly bound. 
Infrared and radio observations of the region reveal no such massive stellar cluster, nor distribution of diffuse stars, from which multiple SN progenitors may have originated, and thus we deem this explanation implausible.

An alternative explanation is that the M0.8$-$0.2 ring is the remnant of a hypernova explosion. 
A hypernova is defined as an event where the initial injected energy is greater than the canonical $10^{51}$\,erg for a SN \citep{Paczynski1988}. 
Hypernovae have been proposed to explain the similarly energetic young remnant NGC 5471B in M101 and the synchrotron superbubble in IC10 \citep{Wang1999,Chen2002,Lozinskaya2007}, as well as the Cygnus superbubble \citep{Kimura2013, Bluem2020}. 
It is important to note, however, the discrepancy in scale when making such comparisons, with NGC 5471B, the IC10 superbubble, and the Cygnus superbubble having significantly larger diameters of approximately 70, 250, and 450 pc, respectively \citep{Chen2002, Lozinskaya2007, Kimura2013}. This discrepancy may arise due to the high gas densities and pressures in the CMZ, serving to constrain the size of the M0.8$-$0.2 ring and resulting in the highly energetic remnant remaining compact.
Given the lack of an observed massive stellar cluster in the immediate vicinity, the progenitor may have been a runaway star of very high mass from a nearby cluster.
The striations visible in the channel maps, with filaments extending from the top and bottom of the ring-like structure towards the Galactic west, suggest that the explosion's source might have been a runaway star originating near Sgr B1. The cavity's axis, as seen in the MeerKAT image, aligns with motion towards the Galactic east. Additionally, the increased brightness of non-thermal emission and dust towards the Galactic east supports the idea that the star's winds and ionising radiation created a tunnel extending from Sgr B1 in the direction of the Galactic east before the explosion.

We also checked for consistency with the typical relationship held by SNRs between radio surface brightness and size ($\Sigma-r$; see e.g. \citealp{Case1998}). 
For a mean radio brightness over the region of $\sim\,0.1$ mJy/beam, measured from the MeerKAT observations at 1.28\,GHz with a beam size of 4\arcsec, the surface brightness at 1 GHz for a spectral index of around $-1$ is $\Sigma_\mathrm{1 GHz}\approx 3\times10^{-21}$ W\,Hz$^{-1}$\,m$^{-2}$\,sr$^{-1}$.
Using the $\Sigma_\mathrm{1 GHz} \propto r^{-2.64}$ relation from \citet{Case1998}, a radius of $r \sim 20$\,pc is predicted for such a surface brightness.
This is inconsistent with our calculation of the radius of the M0.8$-$0.2 ring of $\sim6$\,pc, assuming the region is within the Galactic Centre (i.e. at $\sim$\,8\,kpc).
The deviation from the typical $\Sigma-r$ relation may again originate due to the unique nature of the CMZ and the environment the SN explosion occurred in. SNRs found within dense molecular clouds (so-called `dark' or `buried' SNRs; see \citealp{Sofue2020, Sofue2021, Sofue2024}) have been proposed to obey a different $\Sigma-r$ relation from the typical SNRs located in the inter-cloud ISM. This effect may then be compounded by the order of magnitude higher densities and pressures within the Galactic Centre, resulting in a remnant of unusually compact nature.

\subsection{Possible alternatives to stellar feedback?}
\label{subsec_alternatives}

As stellar feedback does not provide a definitive answer for the cause of the M0.8$-$0.2 ring, we next considered possible alternative explanations. 

\subsubsection{Cloud--cloud collisions}

An alternative scenario responsible for the formation of the M0.8$-$0.2 ring could involve cloud--cloud collisions. 
The CMZ harbours a high density of massive molecular clouds, and given the highly dynamic nature of the region, numerous interactions are anticipated. 
Observable signatures of these collisions, often subject to debate, have been identified in several well-known molecular clouds within the CMZ (e.g. \citealp{Higuchi2014, Johnston2014, Tsuboi2015b, Armijos-Abendano2020}). 
Common arguments involve the presence of shocked gas tracers such as HNCO, SiO, and SO, which indeed all exhibit widespread distribution across the region.
Additional indicators of cloud--cloud collisions include kinematic features, such as the widely discussed broad bridge features \citep{Haworth2015}. 
However, inspecting various PV cuts across the region (Fig. \ref{fig_pv-cuts}) revealed no clear examples of distinct bridge features. 
It is important to note though that the absence of such features in our study does not necessarily negate their presence, and a comprehensive kinematic analysis of the region is planned for future work.

That said, the hole-like structure is intriguing in this context. 
A possible scenario could be that during the collision of the two clouds, more compact cloud punched a hole through the centre of the other, resulting in the ring-like structure that we now observe. 
A similar scenario has been presented to explain the kinematic structure observed in Sgr B2 (e.g. \citealp{Hasegawa1994, Sato2000}). 

The location of the region is particularly favourable for a collisional scenario. 
Firstly, it is close to the point where we expect gas to be flowing along the bar and into the Galactic Centre around this longitude (at around the 1.3 degree cloud; see \citealp{Henshaw2023}). 
Interactions between this inflowing gas and pre-existing gas within the CMZ are commonly seen in simulations (see e.g. \citealp{Tress2020a, Sormani2020, Hatchfield2021}), potentially resulting in high density and highly shocked molecular clouds, akin to our observations. 
It is also close to the apocentre of the CMZ orbit, which could lead to a build-up of material - similar to a traffic jam - and thus to cloud--cloud collisions on the orbit, or with clouds in the inflow from the bar.  
The velocity gradient across the M0.8$-$0.2 ring would be in agreement with this scenario; having the highly redshifted velocities on the left side where gas is infalling towards the CMZ along dust-lanes that are oriented away from us (see the geometry presented in \citealp{Henshaw2023}).
That said, it is noteworthy that the systemic velocity of the ring is notably lower (at $\sim$ 35 \kms) than expected for regions in this location to be interacting with the inflowing gas stream (at $\sim$ 100 \kms). The exact location of the inflowing material also remains a subject of significant debate in the literature. 

\subsubsection{Shear and large-scale processes}

Gravitational shear may also play a role in the formation of the observed structure.
Shear forces due to the rapid rotation around the Galactic Centre are expected to be very strong within the CMZ and play a key role in its evolution (e.g. \citealp{Longmore2013b}). 
The projected galactocentric distance of the centre of the M0.8$-$0.2 ring is $\sim$\,120\,pc (not accounting for the three-dimensional location). 
At this galactocentric radius ($r_\mathrm{gal}$), there is a steep increase in the rotation curve of the galaxy from 100 to 200\,\kms\ (e.g. \citealp{Launhardt2002}), and the dimensionless rate of shear depends sensitively on its derivative as $1 - \beta = 1 - \mathrm{d ln} (v_\phi) / \mathrm{d ln} (r)$ (e.g. \citealp{Krumholz2015}). 
Strong shear forces could then be acting to stretch and enlarge the structure.
The combination of shear and feedback may be a potential solution to the energy problem presented in Sect. \ref{subsec_supernova}. 
We could foresee a scenario whereby a SN initially seeds the expansion of the molecular cloud, which is then (azimuthally) enlarged by shear as it evolves. 
This would also explain the origin of the non-thermal emission towards the region, and could provide a reason for our assumed line-of-sight elongation of the structure.
However, this then raises the question as to whether the calculated age of the M0.8$-$0.2 ring is sufficient for the shear forces to have a significant effect.
Moreover, it is also worth mentioning that the velocity gradient that we observe is in the opposite direction to that of, for example, the Brick, and also in the opposite direction of that of cloud-scale simulations that include the shear (e.g. \citealp{Kruijssen2019a}).

An alternative point worth mentioning is that the shear and large-scale cyclical effects within the CMZ could also impart angular momentum on a molecular cloud (e.g. \citealp{Federrath2016}). 
The observed kinematics could then be interpreted as rotation, producing the smooth velocity gradient across the region of $\Delta v \approx 3$\,\kms\,pc$^{-1}$. 
Indeed, the region is reminiscent of a large-scale vortex or eddy-like structure (Fig.\,\ref{fig_CMZ_HNCO_peak}).
The cause of such a structure, and reason as to why this is seen within the plane of the sky (as opposed to in the plane of the disk), is not clear, but the location of this region towards the outer edge of the CMZ, and close to the inflowing material along the dust lanes may not be coincidental. 
Indeed, there may be some connection here to material that is inflowing but overshooting the Galactic Centre \citep{Hatchfield2021}, where the viscous interaction of the rapid inflowing gas grazing the CMZ produces strongly rotating clouds.
However, since the exact location of these contact points remains unclear, this scenario is still open to interpretation.  

\subsubsection{Line-of-sight coincidence}

Lastly, the most simplistic argument may be that the structures that we identify as the M0.8$-$0.2 ring are in fact a chance alignment of associated velocity components along the line-of-sight. 
The three-dimensional spatial structure of the Galactic Centre, and how that relates to the structure of molecular lines in velocity space (i.e. position-position-position vs position-position-velocity), is indeed a subject of much current debate (e.g. \citealp{Henshaw2023, Hu2023}). 
That said, the structure does appear to be coherent spatially and in velocity across a number of molecular lines (see the channel maps for HNCO, HC$_{3}$N, SO, and SiO in  Figs.\,\ref{fig_HNCO_chans} - \ref{fig_SiO_chans}).
Moreover, \citet{Butterfield2024} identified this structure as coherent in polarisation using the SOFIA/HAWC+ data from the FIREPLACE survey \citep{Butterfield2023, Pare2024}, with the polarisation vectors clearly tracing the ACES HNCO emission, as displayed in Fig.\,\ref{fig_multiwavelength}.
Therefore, it seems highly unlikely that this large-scale ring-like structure is misidentified.

\section{Conclusions}
\label{sec_conclusions}

We have studied a complex, massive molecular structure in the CMZ using data from the ALMA large programme ACES.
We find that the M0.8$-$0.2 ring has a very striking appearance in the new high-spatial dynamic range observations of the HNCO(4$-$3) line, which serves as a tracer of shocked, dense gas.
The structure appears to be coherent in its kinematics across a size-scale of $\sim$\,10\,pc, and radiative transfer modelling of the HNCO emission produces a mass estimate of around $10^6$\,\sol.
For context, this is equivalent to some of the most massive and densest star-forming molecular clouds within the Galactic Centre.  
Previous lower-resolution observations identified this cloud as an expanding ring-like structure (e.g. \citealp{Pierce-Price2000, Tsuboi2015}) originating due to stellar feedback. 
We investigated this, and reached the following conclusions. 
\begin{enumerate}
    \item We argue against the possibility of the M0.8$-$0.2 ring being the result of early-stage stellar feedback. A compact radio point source is observed in the ACES H40$\alpha$ line; however, the source has an inconsistent velocity with the ring-like structure and is deemed to be unassociated. While there remains the possibility that there is a population of highly extincted and/or diffuse stars driving the expansion that we are currently unable to detect, this seems unlikely given the lack of evidence in radio observations.    
    \item Supernova feedback could be a viable cause, and would provide an explanation for the non-thermal radio emission observed towards the region. If the M0.8$-$0.2 ring is indeed an SNR, it is certainly atypical, being highly energetic and yet unusually compact for an SNR of its radio brightness.
    The high kinetic energy of the structure could be explained by multiple, clustered SN explosions; however, no massive stellar cluster from which numerous SN progenitors could have originated is observed, and shear makes several explosions within a concentrated volume unlikely. A particularly energetic hypernova could also provide enough energy to power the structure, and if such an explosion occurred within a very dense molecular cloud, this could explain the morphology and kinematics of the region. 
    \item For completeness, we have discussed a number of alternative scenarios for the formation of the observed structure, such as cloud--cloud collisions, shear, and large-scale processes. None of these scenarios provide a clean explanation for the structure, but we suggest promising avenues for future studies of this intriguing region.
\end{enumerate}

\begin{acknowledgements}

We are grateful to the anonymous referee for their constructive and detailed suggestions, which helped significantly improve the quality of this paper.

This paper makes use of the following ALMA data: ADS/JAO.ALMA$\#$2021.1.00172.L ALMA is a partnership of ESO (representing its member states), NSF (USA) and NINS (Japan), together with NRC (Canada), NSTC and ASIAA (Taiwan), and KASI (Republic of Korea), in cooperation with the Republic of Chile. The Joint ALMA Observatory is operated by ESO, AUI/NRAO and NAOJ.

This project made use of the {\sc Astropy}, {\sc APLpy}, {\sc PySpecKit}, {\sc SpectralRadex} python packages \citep{astropy:2013, astropy:2018, astropy:2022, Robitaille2012, Ginsburg2011,Ginsburg2022, Holdship2021}. 

L.C., I.J-S., and V.M.R. acknowledge funding from grants No. PID2019-105552RB-C41 and PID2022-136814NB-I00 by the Spanish Ministry of Science, Innovation and Universities/State Agency of Research MICIU/AEI/10.13039/501100011033 and by ERDF, UE. 
V.M.R. also acknowledges support from the grant number RYC2020-029387-I funded by MICIU/AEI/10.13039/501100011033 and by "ESF, Investing in your future", and from the Consejo Superior de Investigaciones Cient{\'i}ficas (CSIC) and the Centro de Astrobiolog{\'i}a (CAB) through the project 20225AT015 (Proyectos intramurales especiales del CSIC). 
A.S.-M.\ acknowledges support from the RyC2021-032892-I grant funded by MCIN/AEI/10.13039/501100011033 and by the European Union `Next GenerationEU’/PRTR, as well as the program Unidad de Excelencia María de Maeztu CEX2020-001058-M, and support from the PID2020-117710GB-I00 (MCI-AEI-FEDER, UE).
AG and MGSM acknowledge support from the NSF under grants AAG 2008101, 2206511, and CAREER 2142300.
FHL acknowledges support from the ESO Studentship Programme.
JDH gratefully acknowledges financial support from the Royal Society (University Research Fellowship; URF/R1/221620).
MCS acknowledges financial support from the European Research Council under the ERC Starting Grant ``GalFlow'' (grant 101116226).
JEP was supported by the Max-Planck Society.
 
CB gratefully  acknowledges  funding  from  National  Science  Foundation  under  Award  Nos. 2108938, 2206510, and CAREER 2145689, as well as from the National Aeronautics and Space Administration through the Astrophysics Data Analysis Program under Award ``3-D MC: Mapping Circumnuclear Molecular Clouds from X-ray to Radio,” Grant No. 80NSSC22K1125.

MN and FWX acknowledge funding from the European Union’s Horizon 2020 research and innovation programme under grant agreement No 101004719 (ORP).

J.M.D.K. gratefully acknowledges funding from the European Research Council (ERC) under the European Union's Horizon 2020 research and innovation programme via the ERC Starting Grant MUSTANG (grant agreement number 714907). COOL Research DAO is a Decentralised Autonomous Organisation supporting research in astrophysics aimed at uncovering our cosmic origins. E.A.C.M. gratefully acknowledges funding from the National Science Foundation under Award
No. 2206509. J.W. gratefully acknowledges funding from National Science
Foundation under Award Nos. 2108938 and 2206510.

\end{acknowledgements}

\bibliographystyle{aa}
\bibliography{references}

\begin{appendix}

\section{Observational information}
\label{appendixsec_obs}

\begin{table*}
\caption{Spectral line cube information.}   
\label{tab:spectral-line-info}                                  
\centering                                    
\begin{tabular}{lcccccc}                         
\hline\hline                                  
Molecule & Transition & Rest Frequency ($\nu_\mathrm{rest}$) & Upper Energy Level (E$_\mathrm{up}$) & Critical Density (n$_\mathrm{crit}$) & Beam & RMS\\
& &  [GHz] & [K] & [$10^4$ cm$^{-3}$] & [arcsec] & [K]\\
\hline
SO & $2_{2}$--$1_{1}$ & 86.0940 & 19.3 & ... & 2.21 & 0.08\\
H$^{13}$CN & 1--0 & 86.3399 & 4.1 & 25 & 2.72 & 0.09\\
H$^{13}$CO$^{+}$ & 1--0 & 86.7543 & 4.2 & 4.1 & 2.69 & 0.09\\
SiO & 2--1 & 86.8470 & 6.3 & 10 & 2.69 & 0.09\\
HN$^{13}$C & 1--0 & 87.0909 & 4.2 & ... & 2.69 & 0.08\\
HNCO & 4--3 & 87.9252 & 10.5 & 2.1 & 2.81 & 0.24*\\
CS & 2--1 & 97.9810 & 7.1 & 10 & 2.21 & 0.09\\
HC$_{3}$N & 11--10 & 100.0763 & 28.8 & 16 & 2.17 & 0.07\\
\hline                                       
\end{tabular}
\tablefoot{ This table lists the rest frequencies, upper energy levels (taken from Splatalogue; \citealp{Remijan2007}), and critical densities (taken from Table\,2 of \citealp{Barnes2020a}, or Table\,1 of \citealp{Shirley2015}; both assume a kinetic temperature of 20\,K) of the spectral line transitions, as well as the circularised beam sizes and the RMS values of the corresponding cubes. The table is sorted by increasing frequency. Note the HNCO cube used throughout this study has a velocity resolution of 0.5\,\kms, whereas the remaining cubes have a resolution of 3\,\kms. Consequently, the noise level in the HNCO cube is higher than that observed in the remaining cubes.}
\end{table*}

Rest frequencies, upper energy levels, and critical densities of the studied transitions, as well as the beam sizes and RMS values of their corresponding cubes, are listed in Table \ref{tab:spectral-line-info}.

\section{Radial profiles}
\label{appendixsec_radprof}

\begin{figure*}
    \centering
        \includegraphics[width=\textwidth]{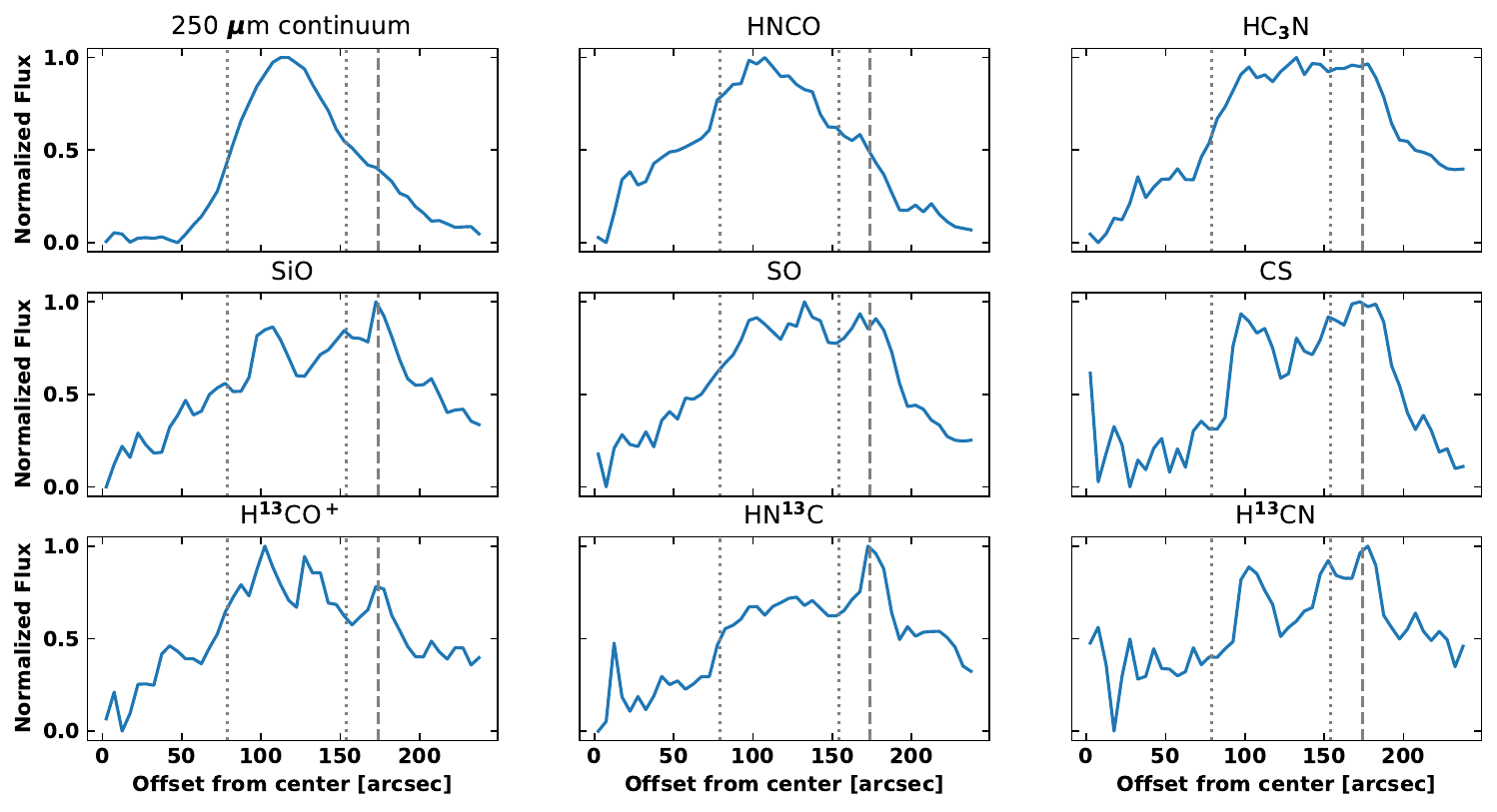}
    \caption{{Radial profiles from ACES spectral lines.} The mean flux contained in annuli of increasing radius is plotted, normalised to the peak value. The superimposed dotted grey lines correspond to the inner and outer radii of the ring-like structure, determined from the continuum emission. The dashed grey line marks another peak at 174$\arcsec$, which is visible in several of the molecular transitions.}
    \label{fig_all_radial}
\end{figure*}

Figure \ref{fig_all_radial} shows the mean flux radial profiles of the spectral lines listed within Table \ref{tab:spectral-line-info} across the M0.8$-$0.2 ring. The annuli are chosen to be 5$\arcsec$ thick and are centred on the central position of the ring-like structure. Most lines show one or two peaks within the continuum-defined extent of the ring (shown by the two dotted vertical lines in Fig. \ref{fig_all_radial}) and another peak at $\sim$174$\arcsec$ (dashed line in Fig. \ref{fig_all_radial}).

\section{Channel maps}
\label{appendixsec_chans}

\begin{figure*}[!tp]
    \centering
        \includegraphics[width=\textwidth]{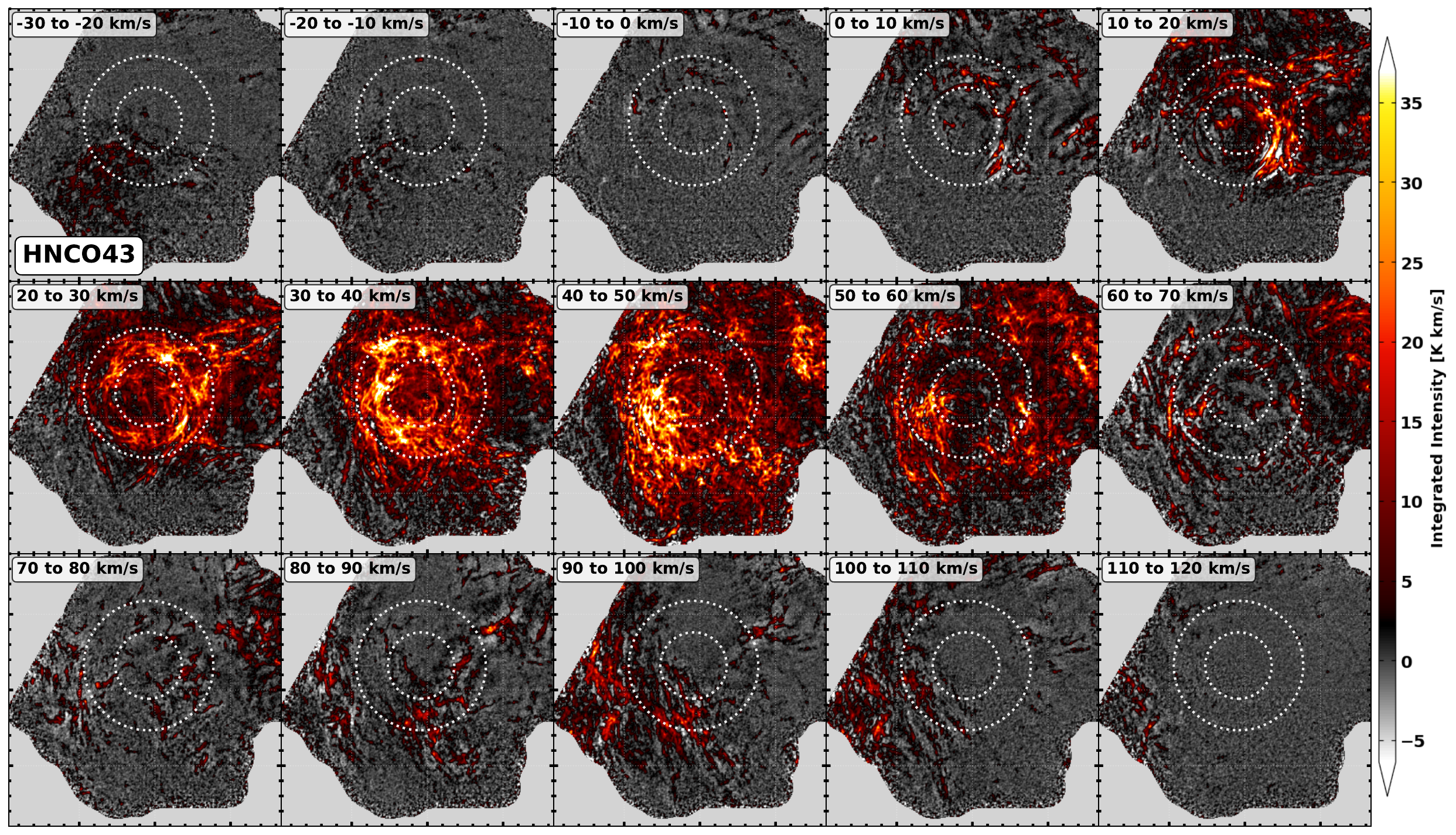}
        \label{fig_chans} \vspace{-3mm}
    \caption{{Channel maps of the HNCO (4--3) emission across the M0.8$-$0.2 ring.} The intensity has been integrated in channels of 10\,\kms\, from $-$30\,\kms\ to 120\,\kms.}
    \label{fig_HNCO_chans}
\end{figure*}

\begin{figure*}[!tp]
    \centering
        \includegraphics[width=\textwidth]{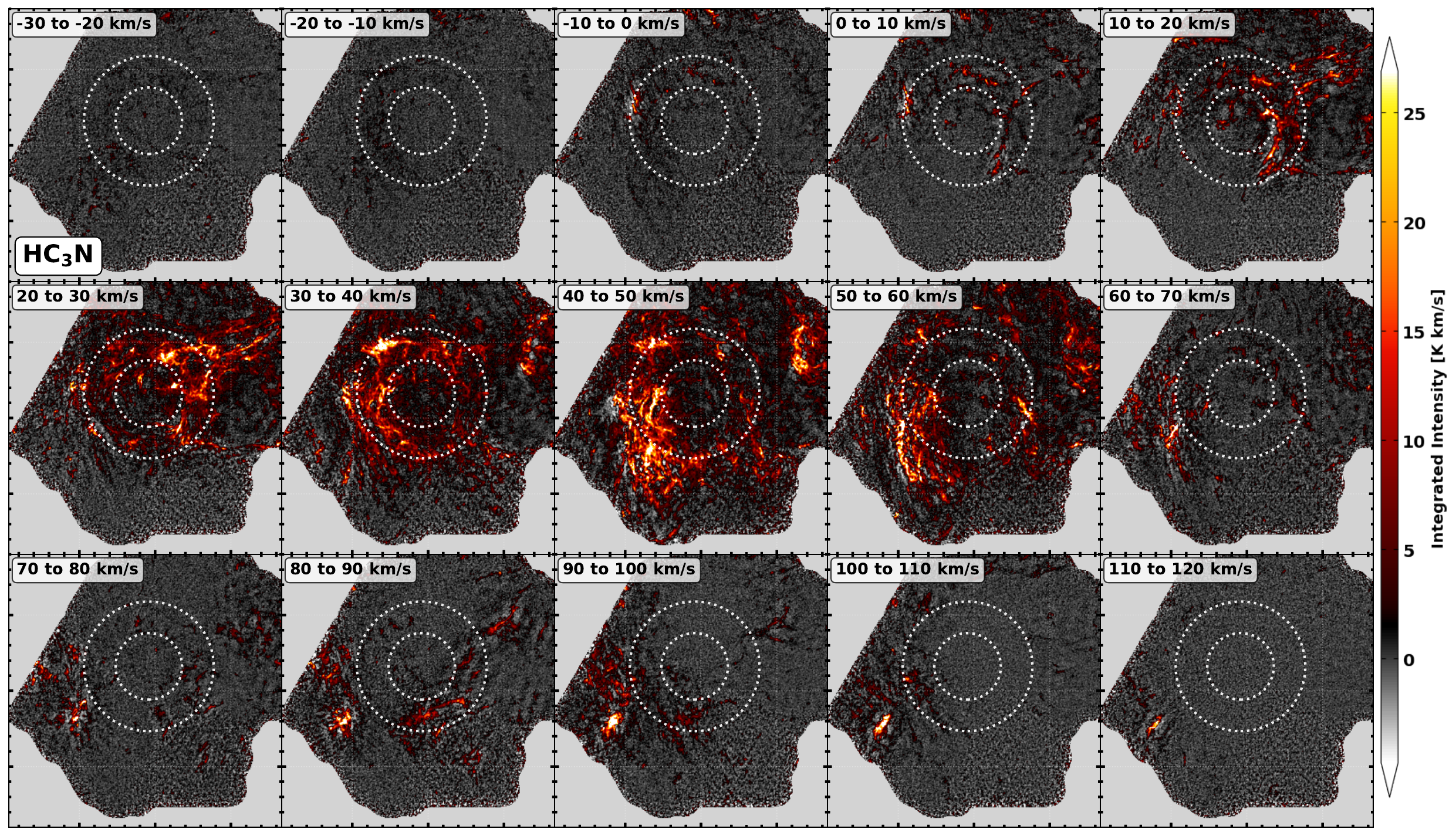}
        \label{fig_chans} \vspace{-3mm}
    \caption{{Channel maps of HC$_{3}$N(11$-$10) emission across the M0.8$-$0.2 ring.} The intensity has been integrated in channels of 10\,\kms\, from $-$30\,\kms\ to 120\,\kms.}
    \label{fig_HC3N_chans}
\end{figure*}

\begin{figure*}[!tp]
    \centering
        \includegraphics[width=\textwidth]{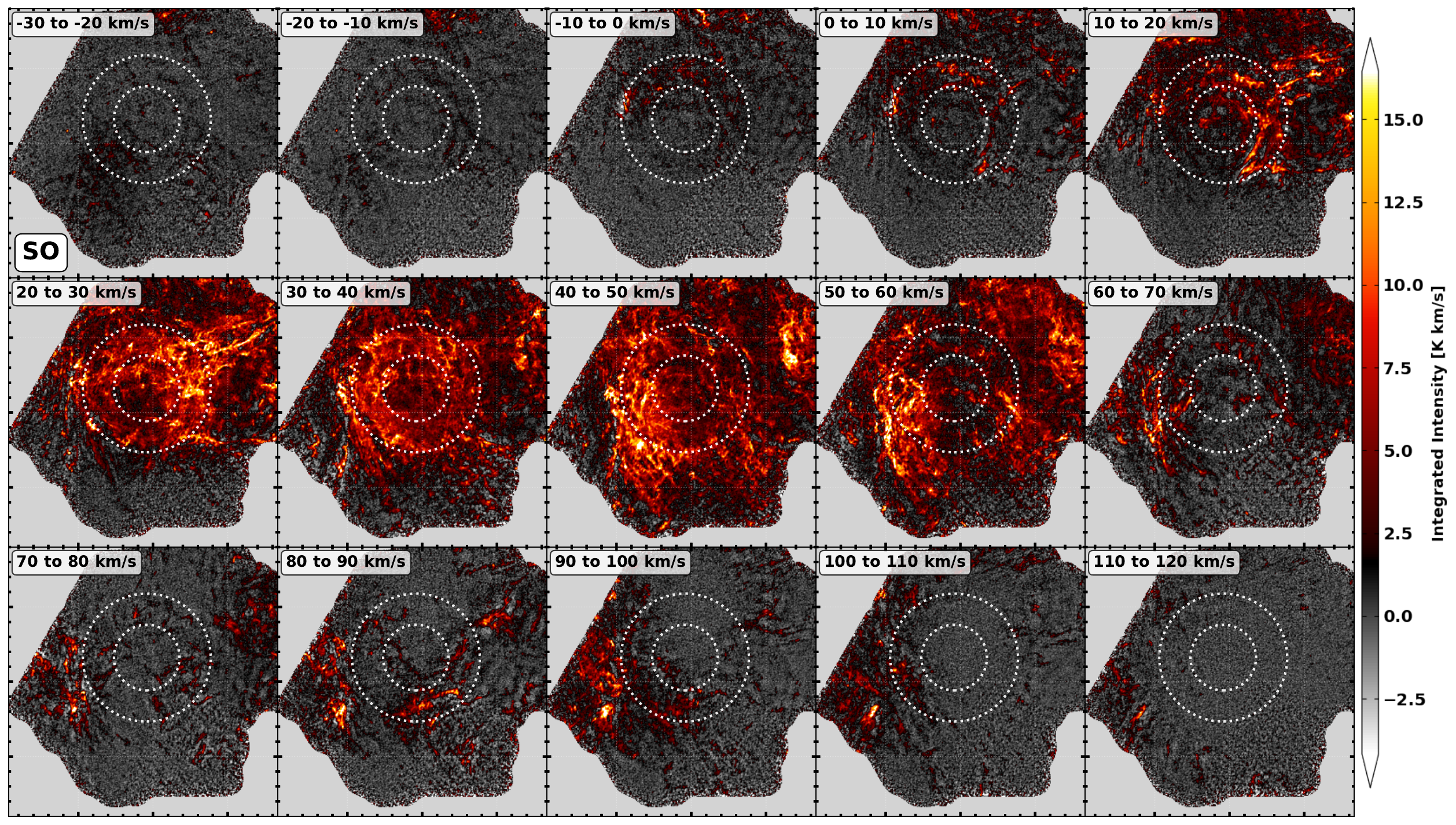}
        \label{fig_chans} \vspace{-3mm}
    \caption{{Channel maps of SO($2_{2}$--$1_{1}$) emission across the M0.8$-$0.2 ring.} The intensity has been integrated in channels of 10\,\kms\, from $-$30\,\kms\ to 120\,\kms.}
    \label{fig_SO_chans}
\end{figure*}

\begin{figure*}[!tp]
    \centering
        \includegraphics[width=\textwidth]{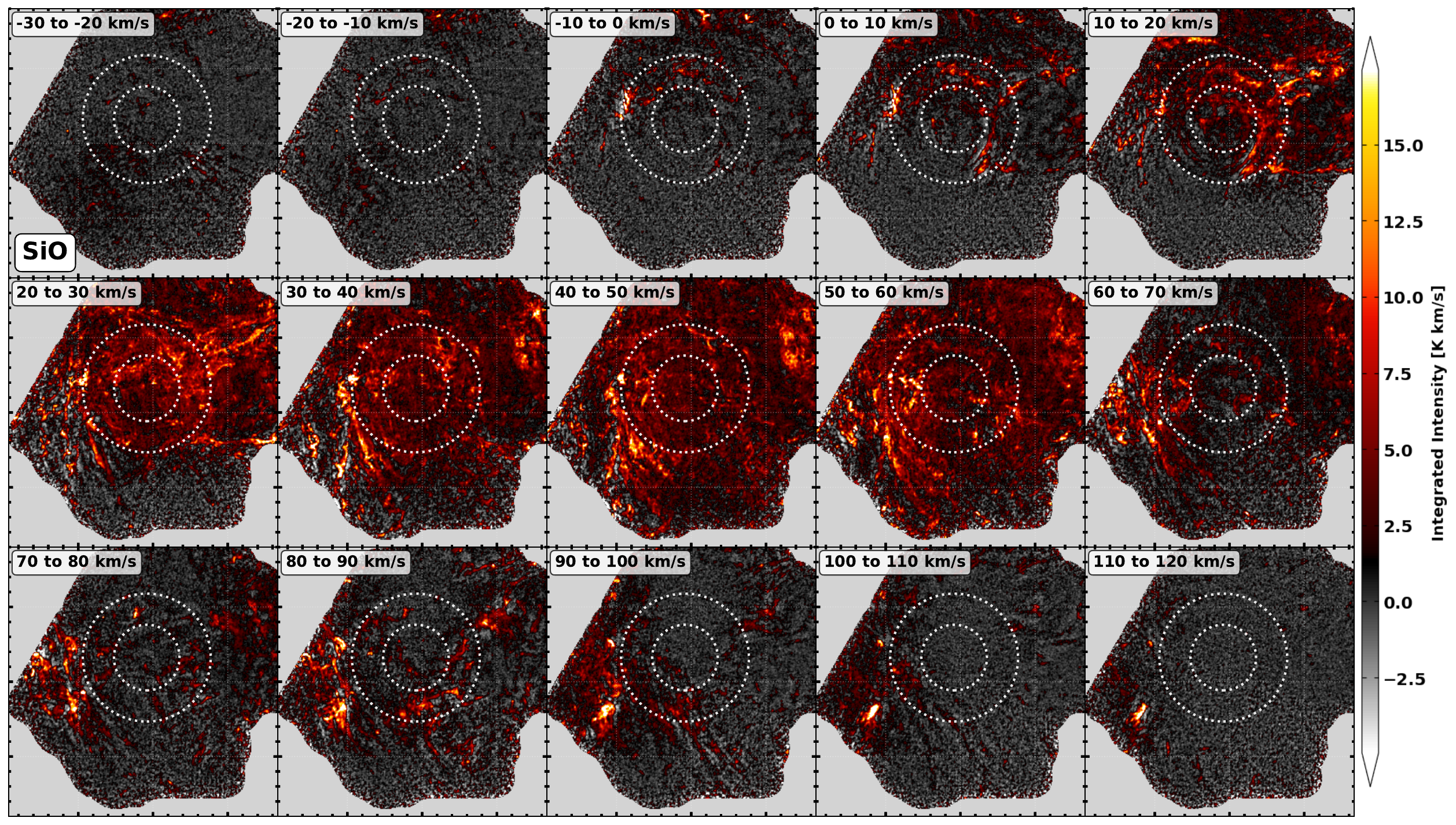}
        \label{fig_chans} \vspace{-3mm}
    \caption{{Channel maps of SiO(2--1) emission across the M0.8$-$0.2 ring.} The intensity has been integrated in channels of 10\,\kms\, from $-$30\,\kms\ to 120\,\kms.}
    \label{fig_SiO_chans}
\end{figure*}

In this section we present channel maps of the HNCO\,(4$-$3), HC$_{3}$N(11$-$10), SO(2$_{2}-$1$_{1}$), and SiO(2$-$1) emission across the M0.8$-$0.2 ring (Figs.\,\ref{fig_HNCO_chans}, \ref{fig_HC3N_chans}, \ref{fig_SO_chans}, and \ref{fig_SiO_chans}, respectively). 
Here, we observe that the ring-like structure is coherent across velocities from 10\,\kms\ to around 70\,\kms\ in the HNCO\,(4--3), HC$_{3}$N(11--10), and SO($2_{2}$--$1_{1}$) transitions.
The SiO(2--1) emission, on the other hand, appears to trace only the outer edge of the structure at similar velocities. 

Above and below the 10\,\kms\ to 70\,\kms\ velocity range, we identify additional velocity components; however, the structure of these components is significantly different from that of the M0.8$-$0.2 ring (Sect. \ref{subsec_morphology}). 
These additional components can also be seen in spectra presented in Figs. \ref{fig_HNCO_spec_moms} and \ref{fig_H40a_spectrum}.
We believe that they are unassociated with the ring-like structure and are instead related to other known velocity features in the Galactic Centre.  
For example, the redshifted (higher) velocity component is consistent with the stream of gas entering the Galactic Centre from the bar (e.g. \citealp{Henshaw2023}). In addition, the redshifted velocity component seen at around 100\,\kms\ is consistent with the emission from the point source (\ion{H}{ii} region) identified in the H40$\alpha$ emission, which is then also connected to this inflowing gas stream.

The channel maps reveal sharp, concentric rings of emission and here we ask how the rings retain their observed sharp structures for 0.1 Myr or more, especially in the highly turbulent interior of the CMZ. 
Magnetic fields may play a role, but unless the magnetic tension and pressures are larger than the turbulent pressure, the rings should be destroyed in a time comparable to the ring thickness divided by the turbulent velocity. 
The individual filaments that make up the rings are observed to have widths comparable to the resolution. 
For widths of a few arcseconds, or $\sim 20,000$ AU $= 0.1$ pc (at the distance of the Galactic Centre), 1 km s$^{-1}$ of turbulence would allow these features to persist for 100,000 years. 
However, at higher values of, for example, 10 km s$^{-1}$, they could only persist for 10,000 years. 
We see that the filaments have velocity dispersions on the scale of the beam size between these two values (see the channel maps in Fig.\,\ref{fig_HNCO_chans}), highlighting that they may be transient features of the region, or pointing to the M0.8-0.2 ring being younger than proposed.

\end{appendix}

\end{document}